\UseRawInputEncoding
\documentclass[twocolumn]{aastex631}

\newcommand{\g}{$\gamma$}
\newcommand{\s}{$s$}

\defcitealias{Sajadian2025}{S25}

\usepackage{graphicx}
\usepackage{hyperref}
\usepackage{amsmath}
\usepackage{booktabs}
\usepackage{ulem}

\usepackage{makecell}
\usepackage[T1]{fontenc}

\begin{document}

\title{Microlensify: a Transformer Based Machine Learning Classifier for Microlensing Events Trained on TESS Light Curves}

\author[0009-0006-2285-6792]{Atousa Kalantari}
\affiliation{Department of Physics, Institute for Advanced Studies in Basic Sciences (IASBS), Zanjan 45137-66731, Iran}
\email{atousakalantari99@gmail.com}

\author[0000-0002-1910-7065]{Somayeh Khakpash}
\affiliation{Department of Physics, Lehigh University, 16 Memorial Drive East, Bethlehem, PA 18015, USA}
\affiliation{Department of Astronomy, University of Virginia, 530 McCormick Road, Charlottesville, VA 22904, USA}
\email{somayeh.khakpash@gmail.com}

\author[0000-0002-2859-1071]{Sedighe Sajadian}
\affiliation{Department of Physics, Isfahan University of Technology, Isfahan 84156-83111, Iran}

\author[0000-0002-9058-9677]{Hosein Haghi}
\affiliation{Department of Physics, Institute for Advanced Studies in Basic Sciences (IASBS), Zanjan 45137-66731, Iran}

\author[0000-0001-7559-7890]{Willow Fox Fortino}
\affiliation{University of Delaware, Department of Physics and Astronomy, 217 Sharp Lab, Newark, DE 19716 USA}

\author[0000-0003-0972-1376]{Rosanne Di Stefano}
\affiliation{Harvard-Smithsonian Center for Astrophysics, Cambridge, MA, 02138, USA}

\begin{abstract}
Microlensing can reveal populations of faint compact objects that are otherwise difficult to detect. Depending on their design, all--sky surveys have the potential to search for these objects across the sky. The Transiting Exoplanet Survey Satellite (TESS), primarily designed to detect transiting exoplanets, also provides near all--sky coverage with high cadence. In this work, we use TESS data to search for microlensing candidates using both traditional and machine-learning methods and to identify associated false positives in high--cadence surveys.
Microlensify is a physics--informed, transformer--based variational autoencoder trained on simulated single--lens microlensing light curves and real TESS Sector 12 data. The model classifies events, reconstructs light curves, and estimates microlensing event durations. Applied to $\sim5.6$~million TESS light curves, it identified between $0.036\%$ and $1.89\%$ as microlensing candidates across different TESS pipelines. After applying microlensing detection metrics and cross-matching with SIMBAD, we obtained a final list of candidates and identified false positives including long--period variables, Mira variables, cataclysmic variables, red giants, and transients. We also found gaussian--like peaks caused by asteroid crossings, a potential source of false positives in high-cadence microlensing surveys.
The model also predicts event duration with an accuracy of $R^2 = 0.97$.
The model was further tested on published events from different ground‑-based microlensing surveys, confirming 92.7\% as microlensing, demonstrating its applicability across surveys with different cadences.
\end{abstract}

\section{Introduction} \label{sec:intro}
Gravitational Microlensing, first described by \citet{Einstein1936}, occurs when the gravitational field of a foreground object (the lens) bends light rays from a background star (the source) along the observer's line of sight. This produces multiple images of the source, separated by only microarcseconds that are smaller than the angular resolution of both ground- and space-based instruments, making them unresolvable.  Instead, the phenomenon is observed as a characteristic brightening and dimming of the source star (\citeauthor{Refsdal1964} \citeyear{Refsdal1964}; \citeauthor{Paczynski1986} \citeyear{Paczynski1986}).

\begin{figure*}[t]
    \centering
    \includegraphics[width=0.8\textwidth]{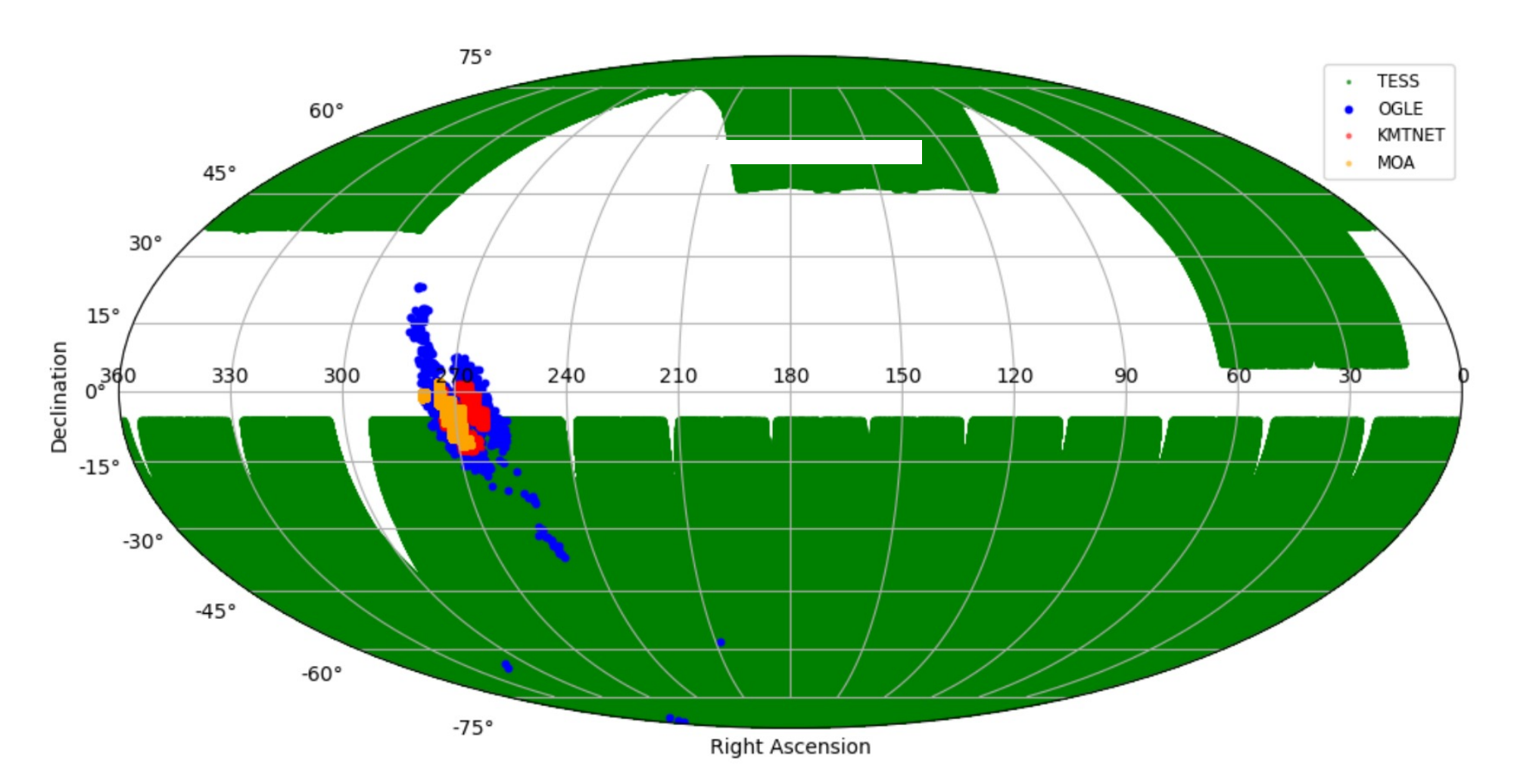} 
    \caption{Distribution of microlensing events from KMTNet, MOA, and OGLE, overlaid with the TESS observing fields in its first two years, plotted in ecliptic coordinates.}
    \label{fig:sky_map}  
\end{figure*}

The detection rate of gravitational microlensing events depends on the stellar density of the observed field; therefore several photometric surveys have been designed to target high-density regions such as the Galactic bulge, the Galactic plane, and the Large and Small Magellanic Clouds (LMC and SMC). Historically, the MACHO \citep{Alcock2000} and EROS \citep{Tisserand2007} projects observed the Galactic halo to search for dark matter candidates using microlensing. Modern ground-based surveys, including the Microlensing Observations in Astrophysics (MOA), the Optical Gravitational Lensing Experiment (OGLE), and the Korea Microlensing Telescope Network (KMTNet), primarily focus on the Galactic bulge to detect exoplanet microlensing events (\citeauthor{Sumi2003} \citeyear{Sumi2003}; \citeauthor{Udalski2015} \citeyear{Udalski2015}; \citeauthor{Kim2016} \citeyear{Kim2016}). Additionally, surveys not specifically designed for microlensing have also detected events, such as the Zwicky Transient Facility (ZTF), which scans the entire northern sky with a 1–-3 day cadence \citep{Rodriguez2022}, and the VISTA Variables in the Vía Láctea (VVV) survey, which maps the Galactic bulge and part of the Galactic plane in the near-infrared \citep{Navarro2020}. The space-based Gaia mission covers the entire sky; on average, each star was scanned about 40 times over the 34-month period used in DR3 \citep{Wyrzykowski2023}.

In addition to these surveys, the Transiting Exoplanet Survey Satellite (TESS\footnote[3]{\url{https://tess.mit.edu/}}) provides a promising, yet relatively unexplored, opportunity for microlensing studies. While TESS was primarily designed to detect exoplanets through the transit method, its unique observing strategy also makes it suitable for detecting short-duration transient phenomena. TESS observes 13 partially overlapping sectors in the southern hemisphere and 13 in the northern hemisphere each year during its primary mission, with each sector covering an area of ${24}^{\circ} \times {96}^{\circ}$. This wide field of view is achieved using four cameras, which continuously take 2-second exposures of the observed region. Each sector is monitored through two continuous 13.7-day orbits around the Earth per lunar orbit, with a short gap of approximately 1-–2 days to transfer the collected data back to Earth \citep{Ricker2015}. The individual 2-second exposures are combined on board to produce several types of data products: Full Frame Images (FFIs), smaller postage-stamp cutouts known as Target Pixel Files (TPFs), and their corresponding light cßurves (LCs).
TPFs are generated for targets selected from the Candidate Target List (CTL), a catalog of bright nearby stars with pre\-measured stellar parameters \citep{Stassun2018,Stassun2019}; during the primary mission, the TPFs and their associated light curves were produced with a 2-minute cadence. The cadence of the FFIs has evolved over the course of the mission, starting at 30 minutes during Cycles 1 and 2, improving to 10 minutes in Cycles 3 and 4, and reaching 200 seconds from Cycle 5 onward \citep{tess_data2025}. Additionally, various pipelines, such as the TESS Science Processing Operations Center (SPOC) pipeline \citep{Caldwell2020}, the GSFC-ELEANOR-LITE \citep{Powell2022}, and the MIT Quick-Look Pipeline (QLP) (\citet{Huang2020a}; \citet{Huang2020b}; \citet{Kunimoto2021}; \citet{Kunimoto2022}), process the FFIs to generate light curves for millions of stars at multiple cadences, whose targets are much larger and less restricted than those in the CTL, providing a rich resource for detecting transient phenomena. In \autoref{fig:sky_map}, the distribution of microlensing events from KMTNet, MOA, and OGLE is shown, covering the Galactic bulge, the Magellanic System, and the Galactic disk, along with TESS observation fields for the first two years of the mission (Sectors 1–-26). 
\citet{Sajadian2025} (hereafter \citetalias{Sajadian2025}) simulated single-lens microlensing events detectable by TESS during the first two years of the mission using Monte Carlo simulations. For CTL stars, they estimated an average optical depth of $\sim 0.2 \times 10^{-9}$ and an event rate of $\Gamma_{\mathrm{TESS}} \sim 0.6 \times 10^{-9}\,\mathrm{star^{-1}\,day^{-1}}$. For stars extracted from the TESS FFIs using the SPOC pipeline, they found higher values, with optical depths of $\sim 1$--$3 \times 10^{-9}$ and event rates of $\Gamma_{\mathrm{TESS}} \sim 1$--$4 \times 10^{-9}\,\mathrm{star^{-1}\,day^{-1}}$, with the highest values occurring in Sector 12. They further estimated that the total expected number of microlensing events for CTL stars is $N_{e,\mathrm{tot}} \sim 0.03$. For FFI stars, the expected number of events per star during a 27.4~day TESS observing sector is $N_{e,\mathrm{tot}} \sim 1.4 \times 10^{-6}$.

To distinguish between microlensing and non-microlensing signals, different selection cuts have been proposed. These include the $\chi^{2}$ of point-source point-lens events, which can be described by a simple analytical formula \citep{Paczynski1986}; the $\Delta\chi^{2}$ from an event-finder algorithm designed for the Korea Microlensing Telescope Network (KMTNet), which performs two linear fits in high- and low-magnification regimes on a grid of single-lens microlensing model parameters \citep{Kim2018}; the skewness-von Neumann parameter space \citep{Rodriguez2022}; and a set of features introduced by \citet{Khakpash2021} for classifying different types of microlensing models and non-microlensing light curves or machine learning algorithms as will be discussed later in this section. Defining cut criteria that balance the number of correct candidates is challenging, and the fitting process is time-consuming when applied to millions of light curves in dedicated surveys. Microlensing events are rare (the first microlensing event was discovered among 1.1 million stars in OGLE; \citealt{Udalski1993}) so a very large number of light curves must be searched. This has led the community to explore faster approaches, particularly those based on machine learning.

\citet{Belokurov2003} used a neural network to distinguish microlensing light curves from other forms of variability.  
\citet{Wyrzykowski2015}; \citet{Wyrzykowski2016} applied a random forest classifier trained on several features of each light curve to identify single-lens microlensing events as well as single-lens events exhibiting a parallax effect due to the Earth's motion around the Sun in the OGLE-III data set. \citet{Chu2019} employed three different machine learning techniques Random Forest, Support Vector Machine, and K-Nearest Neighbor, trained on features extracted from the United Kingdom Infrared Telescope (UKIRT) survey to identify microlensing events, and found that the Random Forest achieved the highest accuracy. \citet{Godines2019} introduced another Random Forest classifier for detecting microlensing events, this time using real-time rather than archival data. Their model was trained on extracted features of simulated light curves of RR~Lyrae, Cepheid variables, cataclysmic variable stars, and microlensing events, achieving 94\% accuracy on simulated data. \citet{Mroz2020} used two simple neural-network algorithms trained on feature-selected OGLE-III and OGLE-IV data to classify single- and binary-lens events, and demonstrated that these models perform well on microlensing events detected in the Zwicky Transient Facility (ZTF).

\citet{Viana2025} developed LensNet, a branched architecture of three recurrent neural networks trained on observation time, flux, sky background, star's FWHM, flux uncertainty, air mass, and point-spread function (PSF) quality. Unlike other methods trained only on extracted features, LensNet focuses on detecting real-time alerts. The training set was constructed from the $\sim$20,000 candidate stars per night that the KMTNet AlertFinder selected for light curve review during the 2021 microlensing season, achieving a classification accuracy above 87.5\%.

Transformers \citep{Vaswani2017}, originally developed for natural language processing, have gained significant success in the deep-learning community due to their ability to model complex patterns in sequential data. In the context of astronomical transients, \citet{Allam2023} applied a novel time-series transformer architecture to the Photometric LSST Astronomical Time-Series Classification Challenge \citep[PLAsTiCC]{Hlozek2020}, which contains imbalanced light curves from various transient classes, including microlensing events. Their transformer achieved a logarithmic loss of 0.507, a micro-averaged receiver operating characteristic area under the curve of 0.98, and a precision–-recall area under the curve of 0.87.
Whereas \citet{Allam2023} trained a transformer in a supervised way, \citet{Aguilar2023} presented ASTROMER, a transformer-based model that creates powerful representations of light curves. ASTROMER was pre-trained in a self-supervised manner on millions of unlabeled R-band MACHO light curves, requiring no human-labeled data. The resulting embeddings can directly enhance the training of classifiers or regressors: when combined with labeled variable stars from MACHO, OGLE-III, and ATLAS, ASTROMER-based models consistently outperform a baseline recurrent neural network trained on raw light curves, achieve state-of-the-art variable star classification results, and require substantially fewer computational resources.

In recent years, physics‑-informed machine learning models have become increasingly popular in astronomy because they incorporate known physical relationships directly into the training process. Rather than relying only on patterns in the data, these models use physical constraints to guide learning and produce more physically meaningful representations. For example, \citet{gagliano2023} used a physics‑-informed Variational Autoencoder to estimate galaxy properties from images, while \citet{andika2025} developed VariLens to reconstruct lensed quasar images and estimate lens parameters.

In this work, we aim to leverage TESS's comprehensive all-sky survey and high cadence to search for single--lens microlensing candidates and to identify microlensing false positives. We introduce Microlensify, which is a Physics-Informed Transformer-Based Variational Autoencoder (PI-TVAE) trained and validated on simulated TESS microlensing light curves and on real GSFC-ELEANOR-LITE pipeline light curves. To further check our microlensing candidate list, we also implement the KMTNet event-finding algorithm together with several traditional detection metrics to evaluate the results. We also test the performance of Microlensify on microlensing candidates from other surveys, including OGLE, MOA, and KMTNet.

The outline of the paper is as follows. In Section~\ref{sec:Microlensing Fundamentals}, we review the fundamentals of gravitational microlensing relevant to this work. Section~\ref{sec:data} describes the data used in this study, including the simulated microlensing events and the TESS light curves used to train Microlensify. In Section~\ref{sec:Methodology}, we present the detection metrics used in our analysis, including the KMTNet event finder algorithm and the signal--to--noise ratio and skewness-–von Neumann parameter space used to identify candidate events. Section~\ref{sec:Machine Learning Model} describes the architecture of Microlensify and the training procedure of the model. In Section~\ref{sec:results}, we present the results of applying Microlensify to TESS data and other survey data, along with the results of cross-matching with Gaia. Finally, Section~\ref{sec:concl} summarizes our findings and presents our conclusions.

\section{Microlensing Fundamentals}\label{sec:Microlensing Fundamentals}

When a foreground star aligns with a background source star within a detector's line of sight, the gravitational field of the foreground star bends the light from the background star, creating multiple images of the source. These images form a characteristic ring, known as the Einstein ring, with a radius called the Einstein radius that is given by:

\begin{equation}
R_E = \sqrt{\frac{4GM_l D_l (1 - D_l / D_s)}{c^2}} 
\end{equation}

where \(M_l\) is the mass of the lens and \(D_l\), \(D_s\) are the lens and source distances. The Einstein crossing time is \(t_E = {R_E}/{v_{rel}} \),
where \(v_{rel}\) is the lens-–source relative velocity.
This phenomenon results in a brightening peak in the source star's observed flux relative to its baseline (unlensed) flux. For a point source and a point lens, this is quantified analytically by the magnification of the source, defined as:

\begin{equation}
\label{eq:magnification}
A(t) = \frac{u^2(t) + 2}{u(t) \sqrt{u^2(t) + 4}}
\end{equation}
where:

\begin{equation}
\label{eq:impact_parameter}
u(t) = \sqrt{u_0^2 + \left(\frac{t - t_0}{t_E}\right)^2} 
\end{equation}

where \( t_0 \) is the time of maximum magnification, \( u_0 \) is the impact parameter (normalized to the Einstein radius). The observed magnified flux at time \( t \) is given by:

\begin{equation}
F(t) = f_s A[u(t; t_0, u_0, t_E)] + f_b
\label{eq:4}
\end{equation}

where \( f_s \) is the source flux and \( f_b \) is any blended background flux.

When the source has a finite size (not point-like) and especially when the impact parameter is comparable to the source size, the lensing magnification factors differ \citet{Witt1994}. The finite source effect on magnification factor is evaluated by the normalized source radius $\rho = R_s D_l / (D_s R_E)$ and generally leading to a reduced peak magnification and a longer event duration.

\section{Data} \label{sec:data}
In this study, we use data from TESS Sector 12, observed during the first year of the mission (May 21–-June 19, 2019). This sector is particularly suitable for microlensing studies, as it covers part of the Galactic bulge and plane where the stellar density is highest. \citetalias{Sajadian2025} also showed that this sector has the highest microlensing detection rate.
We use two sets of data for training: (1) TESS Full-Frame Image (FFI) light curves obtained from publicly available pipelines, including TESS-SPOC, GSFC-ELEANOR-LITE, and QLP; and (2) simulated single-lens microlensing light curves generated to match TESS observational characteristics, as described in the following subsections.

\subsection{Full-Frame Images (FFIs)} \label{subsec:ffis}
The Full-Frame Images (FFIs) are produced by coadding 2-second exposures with different cadences: 30 minutes during Cycles 1 and 2, 10 minutes in Cycles 3 and 4, and 200 seconds from Cycle 5 onward. Sector 12 FFIs were observed with a 30-minute cadence, providing continuous photometric monitoring over about a 27.4--day period, with a gap of 1–-2 days. Different pipelines with distinct methodologies have selected stars in FFIs and produced millions of light curves of them in the corresponding sectors. These pipelines are as follows:

\textit{TESS-SPOC:} TESS-SPOC selects FFI targets through a three-step process: first, all two-minute cadence targets ($\sim$20,000 per sector) are included; second, high-value targets are added, prioritizing stars that are bright in the near-infrared ($H \leq 10$) or nearby ($\leq 100$ pc), with a crowding metric $\geq 0.5$ and TESS magnitude $\leq 16$; third, additional field stars are selected with $\log g \geq 3.5$, crowding $\geq 0.8$, and TESS magnitude $\leq 13.5$ \citep{Caldwell2020}. This approach results in a maximum of $\sim$160,000 targets per sector, making the selection highly selective.

\citetalias{Sajadian2025} used light curves extracted from this pipeline (the only pipeline that includes stellar parameters such as apparent magnitude, blending factor, effective temperature, surface gravity, metallicity, and stellar radius in the header files) to obtain the information needed for simulating single-lens microlensing light curves based on TESS characteristics. We used these simulations to generate our training-set light curves.

In addition, we use the raw flux values from this pipeline, provided as SAP\_FLUX (generated via simple aperture photometry), as input for Microlensify's predictions (see \autoref{sec:results}). Throughout this paper, we refer to this raw flux as SPOC\_Flux.

\textit{QLP:} The QLP selection criteria include stars brighter than 13.5 magnitudes, while stars with TESS magnitudes between 13.5 and 15 are required to have proper motions larger than 200 mas yr$^{-1}$. 
The QLP pipeline performs normalization individually for each 13.7-day spacecraft orbit. As a result, raw flux values are not preserved, and the two orbits within each 27.4-day observing sector are normalized independently.
For our analysis, we use the SAP\_FLUX light curves, which are normalized simple aperture photometry (SAP) light curves extracted from the best aperture \citep{Huang2020a, Huang2020b, Kunimoto2021, Kunimoto2022}. These light curves serve as the input for our prediction process, as will be described in \autoref{sec:results} . Hereafter, we refer to these flux values as QLP\_Flux.

\textit{GSFC-ELEANOR-LITE:} This pipeline provides light curves for all stars brighter than 16th magnitude, creating the largest available TESS FFI dataset, with over 150 million stars. Approximately $6\%$ of faint stars (Tmag $> 14$) per sector may suffer from significant contamination by nearby bright stars \citep{Feinstein2019, Powell2022}. We used Eleanor's principal component analysis (PCA)-corrected flux (PCA\_FLUX), which removes shared systematics through principal component analysis on cotrending basis vectors from the SPOC pipeline to generate flat light curves for injecting synthetic microlensing events. Hereafter, we refer to this flux as Eleanor\_PCA\_Flux.
We also used RAW\_FLUX (obtained by multiplying the aperture and the target pixel file (TPF) and summing the pixel product in each cadence, after background subtraction) to provide non-microlensing examples for training and testing Microlensify. Hereafter, we refer to this flux as Eleanor\_Flux.

Despite the contamination risk in crowded fields, Eleanor is particularly valuable for our work because, unlike QLP, it provides raw flux rather than normalized flux, which is essential for accurate simulations. Unlike TESS-SPOC, Eleanor applies minimal selection criteria, producing a much larger set of light curves and enabling analysis of a significantly broader range of stars.

For all SPOC\_Flux, QLP\_Flux, Eleanor\_Flux, and Eleanor\_PCA\_Flux, we applied a QUALITY = 0 quality-flag filter and removed data points with negative flux values.

\subsection{Simulated Microlensing Light Curves for TESS} \label{subsec:simulations}
To create a realistic set of TESS signals for our detection model, we simulate a large population of single-lens microlensing events based on the characteristics of TESS Sector 12. This process involves generating a theoretical magnification model for each event using Monte Carlo simulations, which was performed in \citetalias{Sajadian2025}; we provide a summary here and refer the reader to that paper for a complete description. Next, we need to generate flattened TESS light curves to inject the theoretical model into them and also account for TESS-like photometric error.

\subsubsection{Monte Carlo Simulations} \label{subsubsec:MC}
In this subsection, we explain how we set up priors on simulation parameters. The magnification of a single-lens event is defined by four key parameters: $t_0$, $t_E$, $u_0$ as given in \autoref{eq:magnification} and \autoref{eq:impact_parameter}, and the finite source size $\rho$. $t_0$ is chosen uniformly from the TESS Sector 12 observation window of 27.4~days, accounting for its 1--2 day data gap and the lens impact parameter $u_0$ is chosen smoothly from the range $[0,1]$. 
For $t_E$ as described in \autoref{sec:Microlensing Fundamentals}, we need the Einstein radius $R_E$ and the $v_{rel}$. $R_E$ depends on the lens distance $D_l$, source distance $D_s$, and source radius $R_s$ for which we will explain how the priors are chosen.

The simulation begins with the catalog of 133,329 FFI stars for a given sector from the TESS--SPOC pipeline. While this catalog provides physical properties (including stellar apparent magnitude, blending factor, effective temperature, surface gravity, metallicity, and stellar radius), the distances are not included. To estimate the source distance ($D_s$) for TESS FFI stars, \citetalias{Sajadian2025} matches each source star's physical parameters to the most similar star in the Besan\c{c}on
 Galactic model\footnote[4]{\url{https://model.obs-besancon.fr/}} \citep{Robin2003,Robin2012}, adopting its absolute magnitudes in the \emph{Gaia} $G$, $G_{\mathrm{BP}}$, and $G_{\mathrm{RP}}$ bands. These are converted to the TESS T--band using relations from \citet{Stassun2018,Stassun2019}. To determine their distances, \citetalias{Sajadian2025} accounts for extinction versus distance using the 3D extinction map by \citet{Marshall2006} and the Besan\c{c}on model. Finally, $D_s$ is calculated using the stars' absolute and apparent magnitudes and the extinction versus distance.
For estimating lens mass ($M_l$), range of $M_l \in [0.1\,M_\oplus, 2\,M_\odot]$ from the normalized mass function provided by \citet{Sumi2023} is used. To determine the lens distance ($D_l$) from the observer, \citetalias{Sajadian2025} uses the microlensing event rate function
\(
\frac{d\Gamma}{dD_l} \propto \rho_t(D_l) \, \sqrt{D_l (D_s - D_l) / D_s}
\)
and based on the total mass density at the location of the lens ($\rho_t(D_l)$), $D_l$ is determined.
To calculate the relative velocity between lens and source stars ($v_{rel}$), \citetalias{Sajadian2025} assumes two types of motion for each star: global velocity and dispersion velocity. The process involves obtaining the velocity components for both stars in the observer's coordinate system, projecting the source's velocity onto the plane of the lens, and then subtracting the two vectors to find their relative motion. The components for the dispersion velocity are derived from the Besan\c{c}on model, and their specific values depend on the star's age and its location within the galaxy's structure.

With the full set of physical parameters generated, the final event parameters can be calculated. The Einstein timescale ($t_E$) is determined from the Einstein radius (which depends on $M_l$, $D_l$, and $D_s$) and the relative velocity ($V_\mathrm{rel}$). The finite source size ($\rho$) is calculated by dividing the source radius ($R_s$, from the TESS-SPOC catalog) by the Einstein radius. This parameter is crucial for modeling the finite-source effect, which we incorporate using the VBBINARYLENSING\footnote[5]{\url{http://www.fisica.unisa.it/gravitationastrophysics/VBBinaryLensing.htm}} package \citep{Bozza2010,Bozza2018}.

\subsubsection{Flattened TESS Light Curves}\label{subsubsec:flattening}

To generate flat light curves, with the goal of obtaining at least two flat light curves for each baseline magnitude, we analyzed approximately 500{,}000 light curves extracted from Full Frame Images (FFIs) using the Eleanor pipeline. For each light curve, we converted the Eleanor\_PCA\_Flux to magnitudes using a zeropoint of 20.44 and calculated the median magnitude as baseline magnitude. We then computed the $\chi^2$ of a linear fit to the Eleanor\_PCA\_Flux and selected light curves with the lowest $\chi^2$ values to ensure flatness. To remove outliers in each light curve without changing the cadence, we replaced all data point with a Eleanor\_PCA\_Flux value deviating by more than 2 standard deviations from the light curve's mean with the median Eleanor\_PCA\_Flux of that light curve simillar to sigma clipping but avoiding removal of the data points.
For light curves with the brightest baseline magnitudes, where the $\chi^2$ values of linear fits are larger, we applied an additional method, which we call window repeating. We defined a window size of half, one-third, or one-quarter of the light curve’s length, selecting smaller windows for light curves with larger $\chi^2$ values. This window was slid across the light curve to identify the segment with the lowest standard deviation. We then repeated this low-variability segment to construct a flat baseline across the entire light curve. Finally, we removed outliers from the repeated light curve and selected those with the lowest $\chi^2$ values from the linear fit to ensure flatness.

Next, we matched the baseline magnitude of flat light curves with that of the simulated microlensing models to find at least two flat light curves for each microlensing model.
For each set of parameters (each matched with two different flat light curves), we assigned two blending values from a normal distribution with mean 0.5 and standard deviation 0.3. These blending values were selected such that one falls in the range $(0.2, 0.5)$ and the other in the range $(0.5, 0.8)$. This choice is motivated by the typical TESS pixel size of 21 arcseconds, which often results in significant blending and also the contamination risk in crowded fields of the Eleanor pipeline.

\begin{figure}[htbp]
    \centering
    \includegraphics[width=\columnwidth]{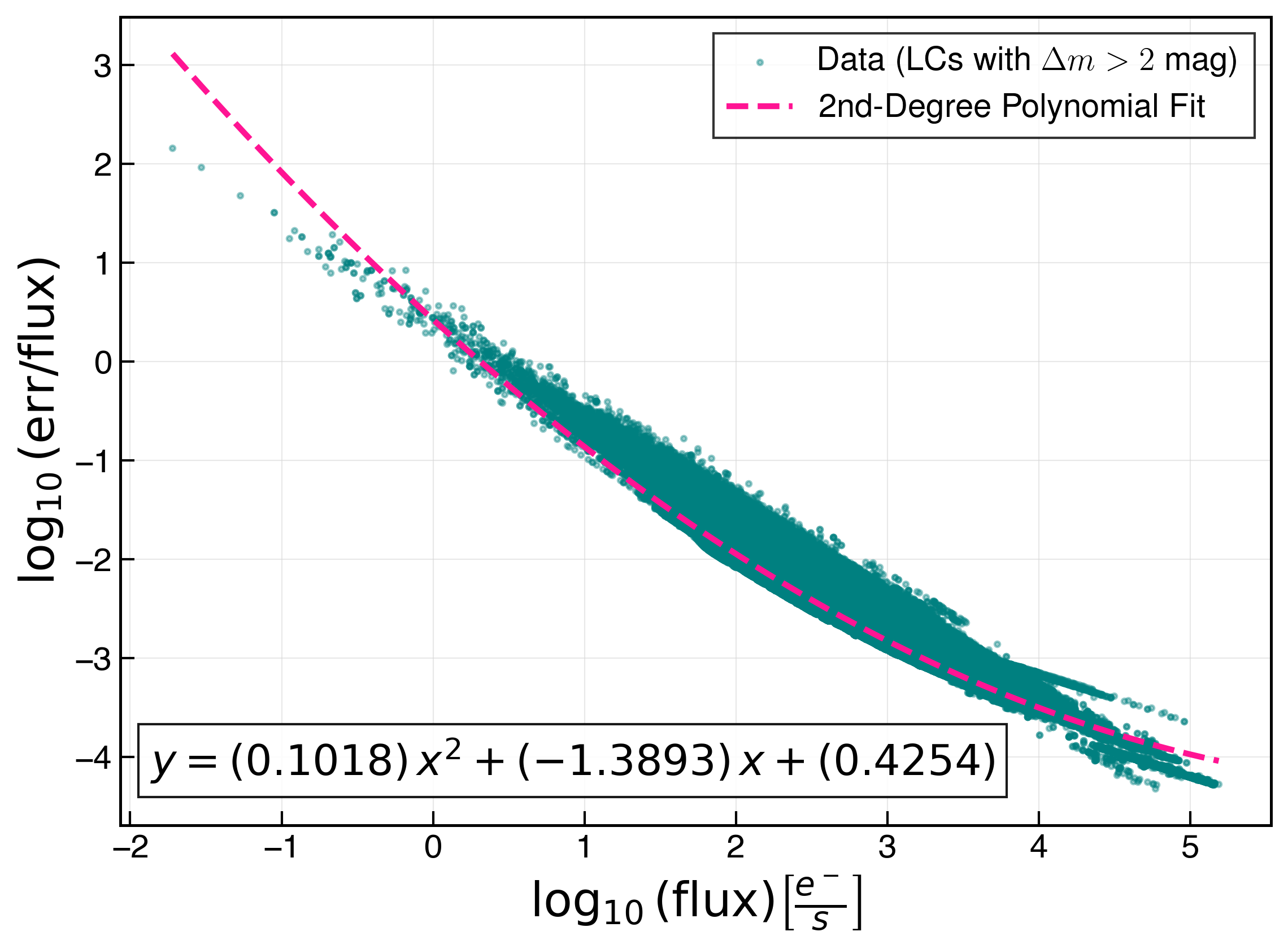}
    \caption{Relationship between $\log_{10}(\mathrm{err}/\mathrm{flux})$ and $\log_{10}(\mathrm{flux})$ for all data points from the 12,000 selected light curves extracted with the Eleanor pipeline, shown as green dots. The pink dashed curve shows the fitted two-dimensional polynomial function.}
    \label{fig:error_function_plot}
\end{figure}


\subsubsection{Experimental flux error function}\label{subsubsec:simulation_error}

We expect that the flux error changes when the source is magnified. To account for this, we select 12,000 TESS light curves with a maximum--minimum magnitude difference greater than 2. For these light curves, we found an experimental relation between flux errors and fluxes.
\autoref{fig:error_function_plot} shows flux error to flux ratio versus fluxes for these
selected TESS light curves along with a polynomial fit. We use this polynomial to change flux errors of magnified data points.

\subsubsection{Injecting Theoretical Model into Flat Light Curve and Applying Error Function}\label{subsubsec:inject}

We simulate the microlensing light curve using the theoretical parameters $t_E$, $u_0$, $\rho$, blending, and $t_0$, where $t_0$ corresponds to the TESS observing time for Sector 12, including its observational gap. Finite-source magnifications are computed with VBBINARYLENSING package for different $u$ and $\rho$ values.

We use the TESS baseline magnitude to generate the theoretical microlensing curve. We then inject that into the TESS light curves that we have selected for baseline creation. This way, we are able to preserve the original scatter in the light curve and the long--term correlated noise of the raw flux.

\begin{figure*}[htbp]
    \centering
    \includegraphics[width=1\textwidth]{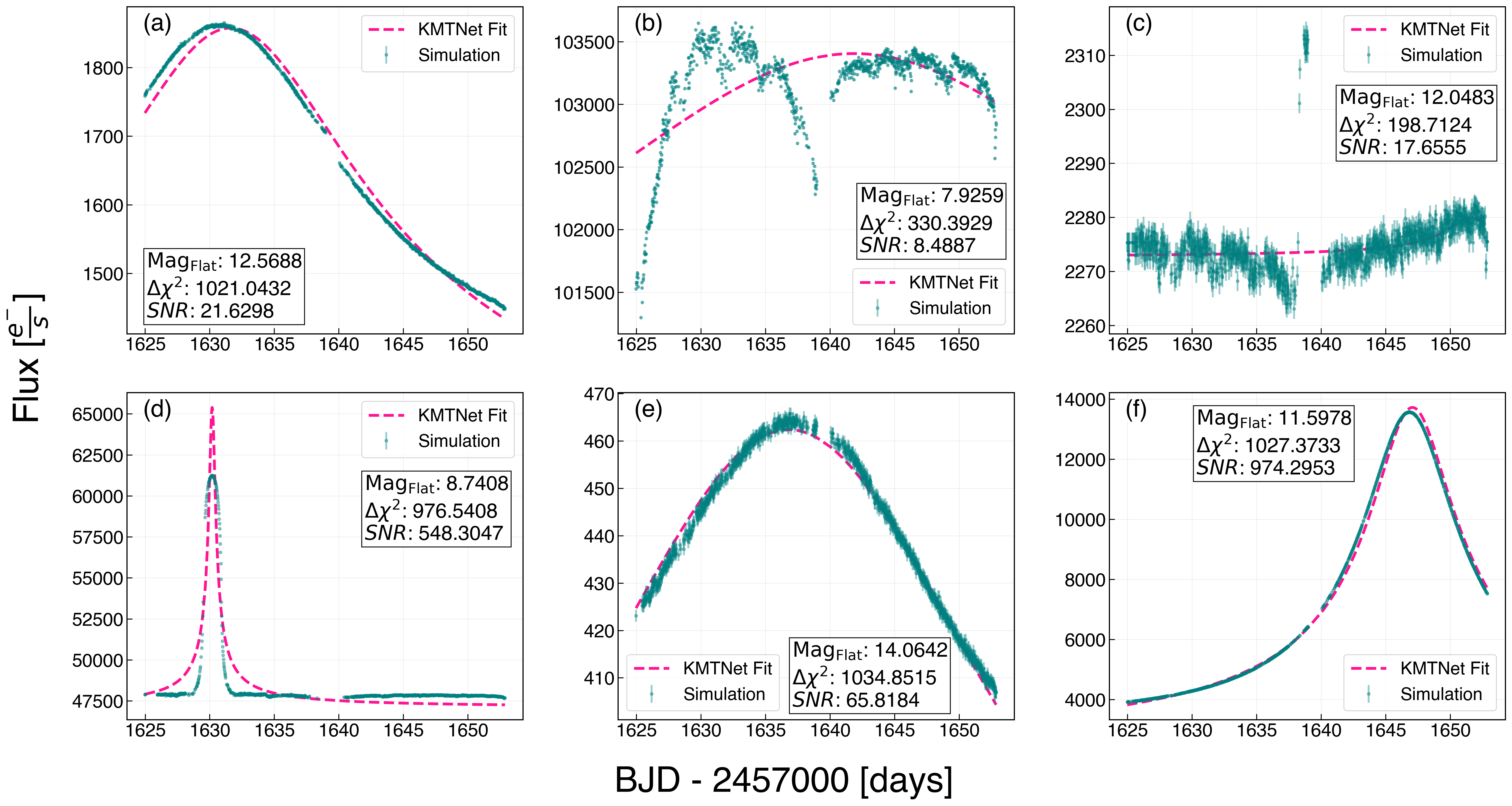}  
    \caption{Simulated TESS microlensing light curves with the corresponding best-fit KMTNet PSPL models overplotted as pink dashed lines. Panels (a)--(c) are rejected by our quality cuts: (a) due to insufficient baseline coverage on one side of the event and low SNR; (b) and (c) due to low SNR, low $\Delta\chi^2_{\mathrm{KMTNet}}$. Panels (d)--(f) satisfy all selection criteria.}
    \label{fig:simulation_kmt}
\end{figure*}

\section{Methodology} \label{sec:Methodology}
We apply a set of selection cuts to construct the training set for our machine learning model, consisting of both simulated and TESS light curves, to ensure that the selected samples represent high-confidence microlensing and non-microlensing candidates. These cuts also help to validate candidates predicted by the model.

\subsection{KMTNet EventFinder Algorithm}
Microlensing events are described by five parameters, as shown in \autoref{eq:4}:
two linear parameters ($f_s$ and $f_b$) and three nonlinear parameters
($t_0$, $t_E$, and $u_0$).

We adopt a fitting algorithm developed by \citet{Kim2018} by rewriting \autoref{eq:4} in two limiting cases:
$u_0 \rightarrow 0$ (high magnification) and $u_0 = 1$ (low magnification),
corresponding to $j=1,2$, respectively.
In these two cases, the observed flux is approximated as
\begin{equation}
F(t) = f_1 A_j\bigl[ Q(t; t_0, t_{\mathrm{eff}}) \bigr] + f_0 , \qquad (j=1,2)
\label{eq:eq6}
\end{equation}

\begin{equation}
Q(t; t_0, t_{\mathrm{eff}}) \equiv 1 + \left( \frac{t - t_0}{t_{\mathrm{eff}}} \right)^2
\end{equation}
where
\begin{equation}
A_1 = Q^{-1/2}, \qquad  (u_0 \to 0); 
\end{equation}
\begin{equation}
A_2 = \frac{Q + 2}{Q \sqrt{Q + 4}}, \qquad (u_0 = 1).
\end{equation}

In the high-magnification limit ($u_0 \rightarrow 0$), $f_1$ approximates the peak flux $F_{\mathrm{max}}$,
while in the $u_0 = 1$ case, $(f_1,f_0) = (f_s,f_b)$.
However, these flux parameters are not identifiable with specific physical quantities in the general case.

After applying these approximations, the model depends only on two nonlinear parameters:
$t_0$ and the effective timescale $t_{\mathrm{eff}} \equiv u_0 t_E$,
resulting in a 2D grid.
The grid points are chosen according to the following sequences:
\begin{equation}
t_{\mathrm{eff},k+1} = \bigl( 1 + \delta t_{\mathrm{eff}} \bigr) t_{\mathrm{eff},k} ,
\end{equation}
\begin{equation}
t_{0,k,l+1} = t_{0,k,l} + \delta t_0 \, t_{\mathrm{eff},k} .
\end{equation}

In this work, we adopt $\delta t_0 = \delta t_{\mathrm{eff}} = \tfrac{1}{5}$ for cases where $t_{\mathrm{eff}} < 1~\mathrm{day}$, and $\delta t_0 = \delta t_{\mathrm{eff}} = \tfrac{1}{3}$ for larger values of $t_{\mathrm{eff}}$. The grid is initialized with $t_{\mathrm{eff},1} \simeq 0.1~\mathrm{day}$ and extended up to a maximum value of $t_{\mathrm{eff}} \simeq 28~\mathrm{days}$, based on the 30-minute cadence and $\sim$ 28-day observation window of TESS Sector 12.

At each grid point, we fit the  $\chi^2$ for both model cases 
(\(j=1,2\)) separately within the window \(t0 \pm Zt_{\text{eff}}\) where \(Z=5\), and determine the best-fit flux parameters \((f_1, f_0)\) by minimizing it using the \texttt{scipy.optimize.minimize} function with the BFGS method \citep{2020SciPy-NMeth}. The $\chi^2$ is defined as:

\begin{equation}
\chi^2 = \sum_{i=1}^N 
\left(
\frac{F_{\mathrm{obs}}(t_i) - F_{\mathrm{model}}(t_i)}
{\sigma(t_i)}
\right)^2,
\label{eq:eq12}
\end{equation}

where \(F_{\mathrm{obs}}(t_i)\) is the observed flux at time \(t_i\), 
\(\sigma(t_i)\) is the corresponding uncertainty, and 
\(F_{\mathrm{model}}(t_i)\) is the model flux computed from \autoref{eq:eq6}.

Once the optimal parameters \((f_1, f_0)\) are found within this window for each model at each grid point, we compute the total $\chi^2$ over the entire light curve using these parameters. The final best-fit parameters are selected as those corresponding to the lowest total $\chi^2$ value across the entire light curve.

To determine the significance of the microlensing signal, we also fit a flat (constant) model to the same time window used for the best-fit parameters, defined by \(t_{\text{best}}-5t_{\text{eff,best}} < t < t_{\text{best}}+5t_{\text{eff,best}}\). The constant flux model is taken to be the weighted mean of the observed flux values within this window. We then compute the $\chi^2$ value for this flat model using \autoref{eq:eq12}, where the model flux \(F_{\text{model},i}\) is replaced by the constant flux level.

Following \citet{Kim2018}, we define a detection metric:
\begin{equation}
\Delta \chi^2 = \left( \frac{\chi^2_{\mu\text{lens}}}{\chi^2_{\text{flat}}} - 1 \right) N
\end{equation}
where \(\chi^2_{\text{flat}}\) and \(\chi^2_{\mu\text{lens}}\) are the $\chi^2$ values of the flat and microlensing fits, respectively, and \(N\) is the total number of data points in the light curve.

Based on simulations, we set a detection threshold of \(\Delta \chi^2 > 400\) for microlensing candidates.

\subsection{Signal-to-Noise Ratio}\label{subsubsec:SNR}
After applying the selection cuts using \(\Delta \chi^2_{\text{KMTNet}}\), we also need to define a threshold on the SNR from our high-confidence microlensing simulations to further confirm the candidates in the TESS light curves.

For the SNR calculation, we follow the approach of \citet{Kunimoto2025}, where the SNR is given by:

\begin{equation}
\mathrm{SNR} = \frac{\delta}{\sigma}, \qquad 
\sigma = \sqrt{ \frac{\sigma_{w}^{2}}{n} + \sigma_{r}^{2} } ,
\end{equation}

where \(\delta\) is the height of the event, \(\sigma_{w}\) is the white noise in the light curve, \(\sigma_{r}\) is the red noise, \(n\) is the number of data points during the event, and \(\sigma\) is the overall noise over the event duration.

To calculate \(\sigma\), we need the event duration. We estimate this using the \(t_{0}\) parameter from the KMTNet fit to find the maximum and half-maximum magnification, and then measure the full width at half maximum (FWHM). In practice, we fix the duration at 4FWHM (four times the FWHM), as this choice was found to lead to a higher recovery rate.

\textit{$\sigma_w$} is defined as the weighted standard deviation of the observed flux with weights $1/(\mathrm{flat\_flux\_err})^2$, where $\mathrm{flat\_flux\_err}$ is the corresponding error after flattening the light curve (\autoref{subsubsec:flattening}), and computed after masking out the data within one event duration. For long-duration simulations that have no baseline, $\sigma_w$ is calculated as the weighted standard deviation of $\mathrm{flat\_flux}$, using weights $1/(\mathrm{flat\_flux\_err})^2$. For TESS candidate light curves with a baseline extracted from the Eleanor pipeline, $\sigma_w$ is computed as the weighted standard deviation of Eleanor\_Flux, using weights $1/(\mathrm{FLUX\_ERR})^2$ (the flux error of the light curve generated by the Eleanor pipeline) and for those without a baseline, $\sigma_w$ is estimated using the median absolute deviation of Eleanor\_Flux.

Following \citet{Hartman2016}, $\sigma_r$ is given by:

\begin{equation}
\sigma_{r}^2 = \sigma_{\mathrm{bin}}^2 - \sigma_{\mathrm{bin,\,uncorrelated\,noise}}^2
\label{eq:sigma_r}
\end{equation}

where $\sigma_{\mathrm{bin}}$ is the weighted standard deviation of the simulated flux after binning in time with a bin size equal to the event duration. The weights are defined as $1/(\mathrm{simulated\_flux\_err})^2$ (calculated in \autoref{subsubsec:simulation_error}) for each bin, and the overall estimate is obtained as the weighted standard deviation of these binned values, with weights $1/(\mathrm{std\_bins})^2$. The term $\sigma_{\mathrm{bin,\,uncorrelated\,noise}}$ is computed in the same way, but using the flattened light curve, with weights $1/(\mathrm{flat\_flux\_err})^2$.
For cases without a baseline, $\sigma_r$ is taken as the weighted standard deviation of the correlated noise, (defined as the difference between $\mathrm{Eleanor\_Flux}$ from the original light curve used for flattening and its $\mathrm{Eleanor\_PCA\_Flux}$) using weights $1/(\mathrm{FLUX\_ERR})^2$.
For TESS candidate light curves with a baseline, we follow the same approach as in the simulations but for $\sigma_{\mathrm{bin}}$, we use Eleanor\_Flux with weights $1/(\mathrm{FLUX\_ERR})^2$ and for uncorrelated noise, we use Eleanor\_PCA\_Flux with weights $1/(\mathrm{FLUX\_ERR})^2$. For cases without a baseline, $\sigma_r$ is calculated as the weighted standard deviation of the correlated noise, using weights $1/(\mathrm{FLUX\_ERR})^2$. If $\sigma_r < 0$, we set $\sigma_r = 0$.

\textit{$\delta$} is defined as the difference between the weighted mean of all data points over the event duration and the weighted mean of points outside the event (for cases without a baseline, the minimum flux is subtracted).
Based on simulations, we set a threshold of \(SNR > 30\) for microlensing candidates.

\subsection{Skewness-Von Neumann Space}\label{subsubsec:Skewness-Von Neumann Space}
Following \citet{Rodriguez2022}, we calculate the skewness $\gamma$ and Von Neumann statistic $\eta$ for all simulated and TESS light curves. They are defined as:

\begin{equation}
\eta \text{ (von Neumann)} = \sum_{i=2}^{N} \frac{ (x_i - x_{i-1})^2}{(N-1) \hat{\sigma}^2}
\end{equation}
\begin{equation}
\gamma \text{ (skewness)} = \sum_{i=1}^{N} \frac{ (x_i - \bar{x})^3}{N \hat{\sigma}^3}
\end{equation}

where $x_i$ is the raw flux, $N$ is the number of data points, $\bar{x}$ is the mean flux, and $\hat{\sigma}$ is the standard deviation of the $x_i$.

The skewness $\gamma$ quantifies the asymmetry of the light curve relative to its mean, whereas the Von Neumann statistic $\eta$ is sensitive to long-term correlated deviations from a constant baseline.

Using our simulated microlensing events, we construct the skewness--Von Neumann space and fit a cubic polynomial function. The fitted coefficients are
\begin{equation}
\eta(\gamma) = a \gamma^3 + b \gamma^2 + c \gamma + d,
\end{equation}
where
$a = 7.419 \times 10^{-5}$,  
$b = -2.315 \times 10^{-4}$,  
$c = 2.502 \times 10^{-4}$,  
$d = 4.358 \times 10^{-4}$.

A light curve is classified as a microlensing candidate if it lies inside this polynomial boundary in the $\gamma$--$\eta$ plane with a tolerance of $10^{-1}$.

\vspace{2mm} 

Along with the cuts on $\Delta\chi^{2}_{\mathrm{KMTNet}} > 400$,  SNR~$> 30$, and being inside the skewness--von Neumann region with a tolerance of 0.1, we also keep only well-sampled simulated light curves by applying two extra cuts.

We remove any light curve that has poor baseline on at least one side of the peak. Specifically, if the coverage before $t_0$ or after $t_0$ is shorter than both half the FWHM of the event and 10 days, the event is rejected. This cut makes sure we have enough data on both sides of the peak, so events with $t_E$ very much larger than 27.4 days whose full peak is not covered by the survey are thrown out.

We also require at least 50 TESS points inside the interval $t_0 \pm \mathrm{FWHM}/2$. This guarantees that the peak is densely sampled.

These two cuts together ensure that the simulated light curves we keep for the training set are of high quality.
\autoref{fig:simulation_kmt} shows the effect of our quality cuts.
Panel (a) is rejected due to insufficient baseline coverage and a low SNR, panel (b) is rejected due to low SNR, low $\Delta\chi^2_{\mathrm{KMTNet}}$, and large $u_0$ that yield negligible peak magnification relative to the noise and panel (c) is rejected due to low SNR and low $\Delta\chi^2_{\mathrm{KMTNet}}$, caused by the small number of data points near the event peak, as it occurred during a TESS observation gap. In contrast, panels (d), (e), and (f) satisfy all selection criteria.

\section{Machine Learning Model}\label{sec:Machine Learning Model}

Neural networks (NNs), a powerful approach within ML, are inspired by the human brain. The foundational element is the fully connected (dense) layer, where every neuron in one layer connects to every neuron in the next. Each connection has an adjustable weight, while each neuron has an adjustable bias. Neurons apply a non-linear activation function to their combined inputs, enabling the network to learn complex relationships. During training, these weights and biases are optimized using a loss function that quantifies and minimizes the difference between the network's predictions and the true targets (as reviewed in \citet{schmidhuber2015deep}).
Alongside fully connected layers, Convolutional Neural Networks (CNNs) use convolutional layers to capture local spatial features by applying small filters (kernels) that slide across the input data. This process creates feature maps that detect patterns like edges and textures, which has made CNNs the standard for image processing. The architecture typically includes pooling layers, which reduce the dimensionality and improve robustness to small spatial translations (see \citealt{oshea2015introduction} for an introductory overview).

Recurrent Neural Networks (RNNs) are a traditional choice for modeling sequential or time-dependent data. Their key feature is a recurrent connection, where the hidden state from a previous time step is fed back as input to the current step. This feedback loop allows the network to retain information from earlier elements in the sequence, making RNNs suitable for tasks such as natural language processing and time-series analysis. However, RNNs face two main limitations. First, their sequential design limits parallelization, which results in slow training on long sequences. Second, since information must pass through many time steps, RNNs often struggle to capture long-range dependencies effectively (see the introduction by \citet{schmidt2019rnn}). The Transformer architecture, introduced by \citet{Vaswani2017}, addresses both of these challenges. Unlike RNNs, it relies on the self-attention mechanism, which models relationships between sequence elements without processing them sequentially. This allows the model to capture dependencies between any two positions in the sequence in parallel. A self-attention block typically contains multiple attention heads, each computing the relevance of every element in the sequence to all others. In a Transformer, the input sequence is first converted into a continuous vector representation using an embedding layer. Because the model processes all elements simultaneously, positional encodings are added to these embeddings to preserve information about the order of the sequence. The embeddings then pass through attention blocks, which combine multi-head self-attention with residual connections, layer normalization, and position-wise feed-forward (dense) layers. For tasks that require sequence generation, the encoder’s output is fed into a decoder. The decoder uses masked self-attention to prevent access to future positions and cross-attention to incorporate information from the encoder. Given these properties, Transformers are now widely used in both natural language processing models and time-series analysis.

Autoencoders (AEs) are neural networks that consist of an encoder, often built with dense or convolutional layers, which compress input data into a low-dimensional latent vector, and a decoder that aims to reconstruct the original input from this vector. A disadvantage of autoencoders is that their unconstrained latent space can lead to overfitting and poor generalization, as the model may memorize training data instead of learning robust features. To address these limitations, Variational Autoencoders (VAEs) \citep{Kingma2019} introduce a probabilistic approach to the latent space. Instead of mapping an input directly to a fixed latent vector, the encoder outputs the mean and variance of a parametric probability distribution, typically a Gaussian. Then a latent vector is randomly sampled from this distribution. This allows the model to produce different outputs for the same input data, with each sample drawn from the corresponding latent distribution.

A specialized version of the Variational Autoencoder (VAE) is the Physics-Informed VAE (PI-VAE), which integrates prior knowledge of the underlying physical system directly into the training process. In this framework, a physical model or known physical relationships are included as additional constraints in the loss function. This approach guides the latent space to encode physically meaningful information, enabling the network to accurately estimate parameters of interest within the latent representation. Recent studies in astronomy highlight the effectiveness of Physics-Informed VAEs. 
\citet{gagliano2023} applied a convolutional PI-VAE to multi-band galaxy images, predicting redshift, stellar mass, and star-formation rate for roughly 26,000 galaxies. Their approach achieved a speed-up of more than 100 times compared to traditional SED fitting and outperformed self-supervised methods in accuracy. \citet{andika2025} developed “VariLens,” a PI-VAE for strong-lens studies, which can reconstruct lensed quasar images, classify lens candidates, and estimate lensing parameters within milliseconds.

In this work, we introduce a Physics-Informed Transformer-Based Variational Autoencoder (PI--TVAE), named Microlensify, for the analysis of TESS light curves, with equally strong performance on data from other surveys. The multi-head attention mechanism effectively captures relationships between different phases of the light curve (baseline, rise, peak, and decay), while the physics-informed variational autoencoder constrains the latent space to estimate the event duration while reconstructing the light curve.

\begin{figure}[htbp]
    \centering
    \includegraphics[width=\columnwidth]{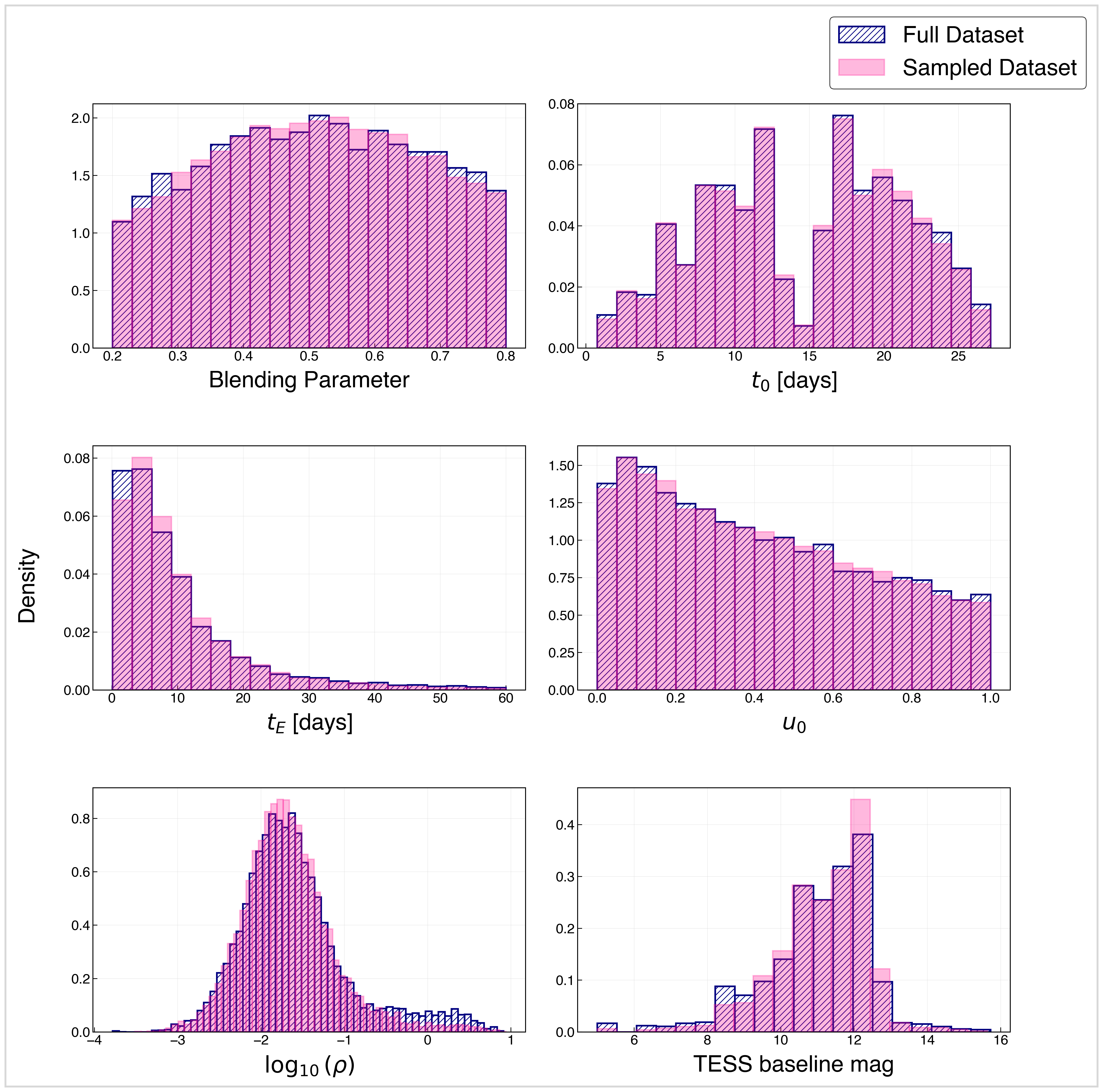}
    \caption{Parameter distributions of the full microlensing simulation sample (blue) and the K-means-selected training subset of 17,000 light curves (pink). Panels show blending ratio, $t_0$ (0-–27.4 days as TESS observation time), $t_E$ (days), $u_0$, $\log_{10}\rho$, and baseline magnitude of the flat light curve. The clustered subset preserves the original distributions.}
    \label{fig:distributions}
\end{figure}

\subsection{Data Preparation for Training}\label{subsec:Data Preparation for Training}

To train Microlensify, we use the flux of simulated microlensing-like light curves together with the Eleanor\_Flux from TESS Sector 12, processed via the Eleanor pipeline, as described in \autoref{sec:data}. For the simulated data, we start with approximately 93,000 light curves that are cleaned with the quality thresholds detailed in \autoref{sec:Methodology}: $\Delta\chi^2_{\mathrm{KMTNet}} > 400$, SNR\,$> 30$, location in the skewness--von~Neumann plane within a tolerance of $10^{-1}$, rejection of events with insufficient baseline coverage on either side of $t_0$ (coverage shorter than both half the FWHM and 10 days), and the requirement of at least 50 TESS points within $t_0 \pm \mathrm{FWHM}/2$. These thresholds ensure that only well-sampled, high-quality simulated events are kept for the training set.

To reduce the computational cost while preserving diversity, we apply K-Means clustering \citep{lloyd1982} to select 17,000 light curves based on six microlensing parameters: blending ratio, $t_0$, $t_E$, $u_0$, $\rho$, and the baseline magnitude of the flat light. \autoref{fig:distributions} shows the histograms of these parameters, confirming that the sampled distributions closely match the original. For the TESS real data, we similarly select 17,000 light curves after removing potential microlensing-like events. Initially, the non-microlensing training set contained 17,000 light curves, and we later added 573 additional examples (Eleanor systematics that had previously been misclassified as microlensing) to improve the model during training. In the final step, we also visually inspect all the light curves to ensure a clean training set. Each light curve, simulated or TESS, is fixed to 940 flux points, corresponding to the average number of points available for TESS Sector 12 after applying a QUALITY = 0 quality-flag filter.

We normalize each light curve individually to the [0, 1] range by applying min-max scaling using its own minimum and maximum flux. Because individual normalization removes differences in amplitude, the model focuses primarily on shape. To enable the model to learn amplitude differences alongside shape for distinguishing microlensing from non-microlensing events, we compute six statistical features per light curve: maximum flux, minimum flux, maximum minus minimum, median absolute deviation, median flux, and standard deviation along with their pairwise ratios. These features are log$_{10}$-transformed and then rescaled to [0, 1]. Before training Microlensify, we train a Random Forest classifier \citep{breiman2001} on the combined simulated (labeled as microlensing) and TESS (labeled as non-microlensing) features to identify the most important features for classification. Features with importance scores greater than 0.1 are retained: maximum, minimum, median, standard deviation, and the ratio of standard deviation to maximum-minus-minimum. These five normalized statistics and the 4FWHM value together with the normalized flux (individually min-max scaled), are fed into the model.
The final dataset comprises 17,000 simulated and 17,573 TESS light curves, each with 940 flux points, five statistics, a 4FWHM value, and binary labels for microlensing classification, split into 85\% training and 15\% validation sets. The Microlensify architecture is detailed in the next section.

\subsection{Microlensify Architecture}\label{subsec:Microlensify Architecture}

In this work, we present Microlensify, a Physics-Informed Transformer-Based Variational Autoencoder (PI-TVAE), for the classification of single-lens microlensing and non-microlensing events, the estimation of the event duration, and the reconstruction the light curves. The model architecture is illustrated in \autoref{fig:architecture}.

The encoder processes two inputs: a light curve with 940 flux points (normalized to [0, 1]) and five statistics (maximum, minimum, median, standard deviation, and the ratio of standard deviation to maximum-minus-minimum), which are re-scaled as detailed in \autoref{subsec:Data Preparation for Training}. The light curve is reshaped to (940, 1), and the five scalar features are expanded to (1, 5) and tiled across the temporal dimension to form a (940, 5) tensor, which is concatenated with the light curve to yield a (940, 6) input.
To capture patterns in the light curve across different time scales, we apply three parallel 1D convolutional layers with kernel sizes of 3, 5, and 7 (filters 64, 128, and 64; Swish activation \citep{ramachandran2017}). These kernels capture short, medium, and long-term patterns in the light curve. The resulting features are concatenated with the input tensor and projected to a target dimension of 64 using a 1D convolutional layer (kernel size 1, Swish activation). This concatenation preserves the original light curve's context alongside extracted features, enhancing the Transformer's input representation.

Sinusoidal positional encodings \citep{Vaswani2017} are added to the sequence embeddings to encode positional information in the sequence, which is essential for light curve sequences. The encoder then applies two transformer blocks, each comprising a multi-head self-attention layer (4 heads, key dimension 16, dropout 0.3), a residual connection, and layer normalization. Unlike standard Transformers that use dense feed-forward networks (FFNs), each block employs two 1D convolutional layers (kernel size 10; filters 16 then 64; Swish activation) with dropout (0.3), residuals and layer normalization, inspired by the Conformer architecture \citep{gulati2020conformer}. The output is then globally flattened, and two additional pre-computed statistical light curve features (skewness and Von-Neumann test statistic, inspired by \citep{Rodriguez2022} as detailed in ~\autoref{subsubsec:Skewness-Von Neumann Space}) are added to the flattened output. Adding these two features consistently improves classification performance over the baseline that omits them.

\begingroup
\setlength{\dbltextfloatsep}{0pt}   
\begin{figure*}[!t]
    \vspace*{-2.5em}               
    \noindent                      
    \makebox[\textwidth][c]{
      \includegraphics[width=\textwidth,height=0.97\textheight,   
                      keepaspectratio]{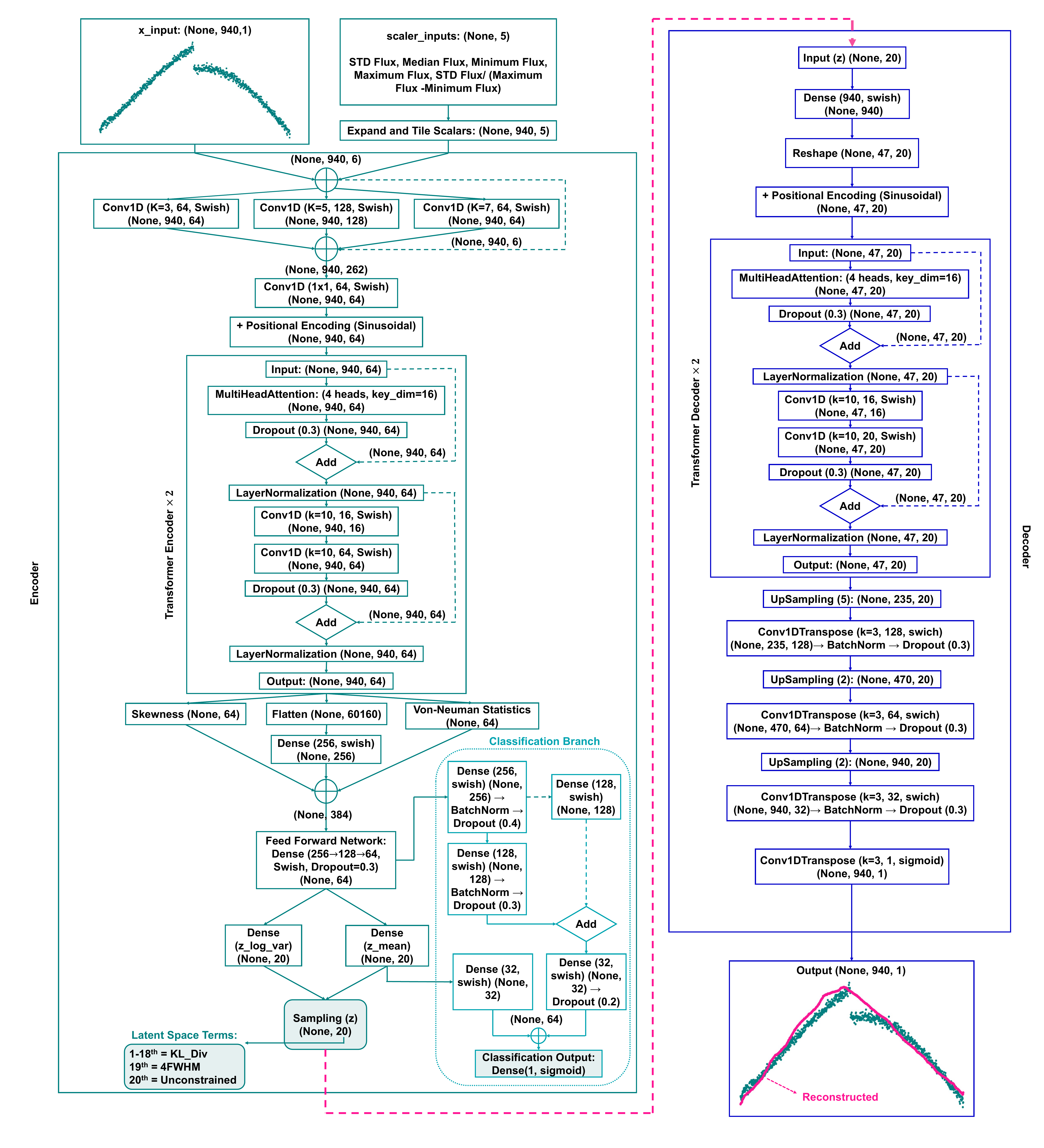}
    }%
    \vfill                         
    \caption{Architecture of Microlensify, a physics-informed transformer-based conditional variational autoencoder trained on $\sim$34{,}500 TESS (Eleanor, Sector 12) light curves. The model takes a normalized flux time series (length 940) and five statistical features as input, simultaneously classifies microlensing vs. non-microlensing candidates, reconstructs the input light curve, and predicts event duration via a physics-informed latent dimension. The example light curve shown in green is the known microlensing event Gaia20fnr (found in TESS with TIC ID 123793270); the model reconstruction is in pink produced by the decoder. For details on this event see \autoref{subsec:Cross-matching Gaia Alerts with TESS}.}
    \label{fig:architecture}
\end{figure*}
\endgroup

\begin{figure*}[!t]
    \centering
    \includegraphics[width=\textwidth]{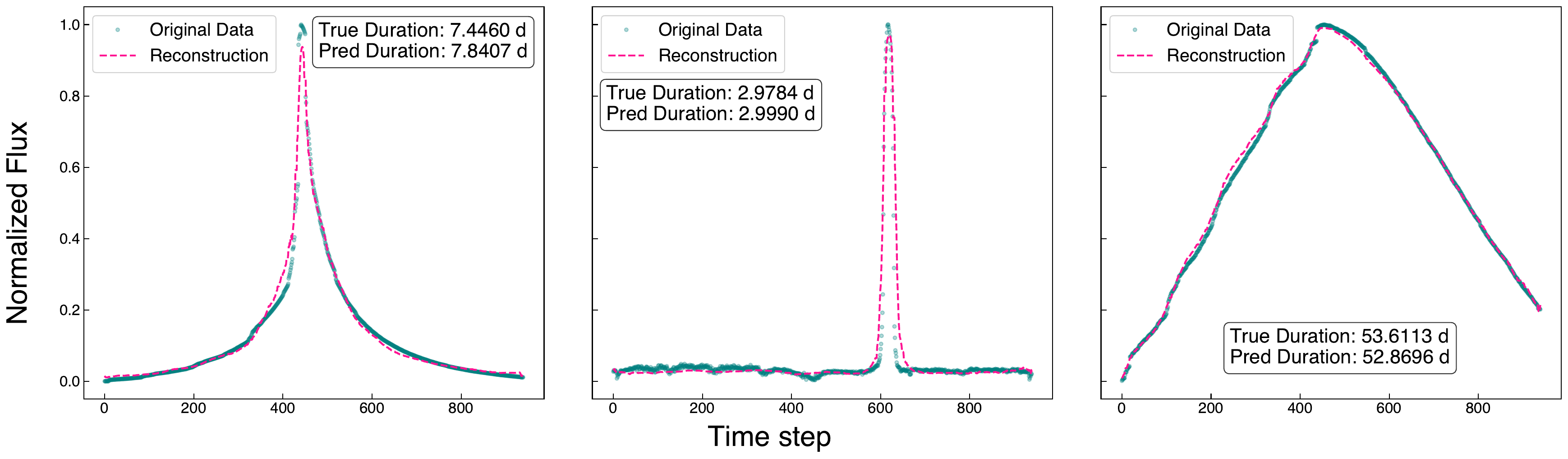}
    \caption{Three true-positive microlensing light curves from the test set. Green points show the original simulation light curve, and the dashed pink line shows Microlensify's reconstruction. For each event, the true duration and the duration predicted by Microlensify, in days, are indicated.}
    \label{fig:true_positives_lightcurves}
\end{figure*}

The flattened features pass through an initial dense layer (256 units, Swish activation), are concatenated with statistical features (skewness and the Von-Neumann test statistic), and are then processed via a feed-forward network (units [256, 128, 64], Swish activation, dropout 0.3) to produce: (1) two separate vectors for the mean ($\mu_z$) and log-variance ($\log(\sigma_z^2)$) of a 20-dimensional latent Gaussian distribution, with 18 dimensions regularized to a standard Gaussian prior via a Kullback-Leibler divergence loss, one dimension constrained to encode the event duration ($4$FWHM) using a physics-informed loss, and one dimension unconstrained; (2) a sampled latent vector $z$ using the reparameterization trick, which separates random sampling from the network's parameters to enable gradient backpropagation, allowing effective learning of the mean and log-variance~\citep{Kingma2019}; and (3) a binary classification prediction for microlensing detection. 

The classification branch processes the output of the feed-forward network through dense layers (256 units with batch normalization and 0.4 dropout, 128 units with batch normalization and 0.3 dropout, all with Swish activation), incorporating a residual connection where the 256 unit output, after batch normalization and 0.4 dropout, is projected to 128 units and added to the 128 unit layer output. The output is then processed through a subsequent 32 unit dense layer with Swish activation, and this is concatenated with the 20-dimensional latent mean projected to 32 units, yielding a 64-dimensional representation that is mapped to a probability for microlensing detection via a final dense layer with sigmoid activation.

The decoder reconstructs the light curve from the 20-dimensional sampled latent vector $z$, without access to the original input or encoder features, following a variational autoencoder framework enhanced with transformer-based sequence processing. A dense layer expands $z$ to a 47 $\times$ 20 tensor (Swish activation), reshaped to (47, 20), with sinusoidal positional encodings. Two transformer blocks, matching the encoder's structure (4 heads, key dimension 16, dropout 0.3), process this sequence using conv1D-based feed-forward networks (kernel size 10, filters 16 then 20, Swish activation) to capture local patterns. Upsampling via factors [5, 2, 2] and 1D transposed convolutional layers (filters [128, 64, 32], kernel size 3, Swish activation, batch normalization, dropout 0.3) restores the sequence length to 940 points. A final 1D transposed convolutional layer (kernel size 3, sigmoid activation) outputs the reconstructed light curve (940, 1).

The Microlensify model is optimized by minimizing a loss consisting of four terms: a reconstruction loss ($\mathcal{L}_{\text{rec}}$), a Kullback-Leibler divergence (KL) loss ($\mathcal{L}_{\text{KL}}$) on 18 out of 20 latent space parameters, a physics-informed KL loss on one out of 20 latent space parameters, and finally a classification loss, as shown in \autoref{eq:tot_loss}.

\begin{align}
\mathcal{L}=\mathcal{L}_{\text{rec}} + \mathcal{L}_{\text{KL,18}} + \mathcal{L}_{\text{KL,dur}} + \mathcal{L}_{\text{class}}
\label{eq:tot_loss}
\end{align}

The reconstruction loss is a weighted sum of a binary cross-entropy (BCE) and a mean squared error (MSE) loss:

\begin{align}
\mathcal{L}_{\text{rec}} &= \alpha \cdot  \sum_{i=1}^N \underbrace{-\left[ x_i \log(\hat{x}_i) + (1 - x_i) \log(1 - \hat{x}_i) \right]}_{\text{BCE}} \notag \\
&\quad + \beta \cdot \sum_{i=1}^N \underbrace{(x_i - \hat{x}_i)^2}_{\text{MSE}},
\label{eq:rec_loss}
\end{align}

where $x_i \in [0,1]$ and $\hat{x}_i$ are the normalized input and reconstructed fluxes, and weights are $\alpha=0.9$, $\beta=0.1$, selected based on training performance.

The Kullback-Leibler divergence loss on 18 out of 20 latent space parameters ($\mathcal{L}_{\text{KL,18}}$) regularizes the latent space, excluding two latent representations intended for the dimension encoding (4FWHM) and an unconstrained dimension:

\begin{equation}
\mathcal{L}_{\text{KL,18}} = \frac{1}{2} \sum_{j=1}^{D-2} \left( 1 + \log(\sigma_{z,j}^2) - \mu_{z,j}^2 - \sigma_{z,j}^2 \right),
\end{equation}
where $\mu_{z,j}$ and $\sigma_{z,j}$ are the mean and standard deviation of the $j$-th latent variable, and $D=20$ \citep{Kingma2019}. This term regularizes the latent space by encouraging the latent variables to follow a Gaussian prior distribution.

The physics-informed KL loss for event duration ($\mathcal{L}_{\text{KL,dur}}$) regularizes the distribution of the dedicated latent dimension to align with a prior centered on the true event duration as in \autoref{eq:dur_loss}. Note that one latent space parameter remains unconstrained, to leave one parameter free to capture any additional feature that might not be as predictable as a known distribution (\citet{gagliano2023}).

\begin{equation}
\mathcal{L}_{\text{KL,dur}} = \log \frac{\sigma_{dur}}{\sigma_{z,dur}} + \frac{\sigma_{z,dur}^2 + (\mu_{z,dur} - \text{dur})^2}{2 \sigma_{dur}^2} - \frac{1}{2},
\label{eq:dur_loss}
\end{equation}

where $\mu_{z,\text{dur}}$ and $\sigma_{z,\text{dur}}$ are the variational mean and standard deviation of the dedicated duration-encoding latent dimension (index 18), $\text{dur} = 4\,\text{FWHM}$ is the true event duration, and the prior standard deviation is fixed at $\sigma_{\text{dur}} = 0.05$ days.

The classification loss uses binary cross-entropy to predict microlensing events:
\begin{equation}
\mathcal{L}_{\text{class}}
= -\left[
y \log(\hat{y}) +
(1 - y)\log(1 - \hat{y})
\right],
\end{equation}
  
where $y \in \{0,1\}$ is the true label and $\hat{y} \in (0,1)$ is the sigmoid-activated output of the classification head, representing the predicted probability of a microlensing event.

\subsection{Training and testing the Model}
\label{subsec:Training}

Microlensify is a Physics-Informed Transformer-Based Variational Autoencoder (PI-TVAE) which is trained for light curve reconstruction, event duration estimation, and microlensing classification. Training is performed using the \texttt{TensorFlow} framework \citep{Abadi2016} on an NVIDIA A100 GPU via Google Colab. The model optimizes a composite loss function consisting of reconstruction loss ($\mathcal{L}_{\text{rec}}$), Kullback-Leibler divergence loss ($\mathcal{L}_{\text{KL}}$), physics-informed duration loss ($\mathcal{L}_{\text{dur}}$), and classification loss ($\mathcal{L}_{\text{class}}$), as defined in \autoref{subsec:Microlensify Architecture}. The dataset of TESS light curves and statistical features, detailed in \autoref{subsec:Data Preparation for Training}, is split into 85\% training data and 15\% validation data. A separate test set of 5500 fully cleaned light curves is used to evaluate the model's ability to achieve zero false positives and zero false negatives in microlensing classification. However, our final evaluation is when we test on uncleaned TESS data to find high--probability microlensing candidates (see \autoref{subsec:Microlensify on sec 12 Eleanor}). A batch size of 64 is used to balance training stability and computational efficiency.

The model was first trained with loss weights: $\mathcal{L}_{\text{rec}}$ at 1.0, $\mathcal{L}_{\text{KL}}$ at 1.0, $\mathcal{L}_{\text{dur}}$ at 1.0, and $\mathcal{L}_{\text{class}}$ at 20.0, empirically tuned to balance reconstruction, latent regularization,  duration estimation, and classification, prioritizing classification for microlensing detection. The model was trained using the AdamW optimizer \citep{Loshchilov2019} with an initial learning rate of $5\!\times\!10^{-4}$, weight decay $10^{-4}$, and global gradient clipping using a clip norm of 0.5.
Training was conducted for up to 200 epochs, with a learning rate scheduler (ReduceLROnPlateau) reducing the learning rate by a factor of 0.5 if the validation total loss plateaued for 15 epochs, with a minimum learning rate of $10^{-6}$. Early stopping halted training if the validation total loss did not decrease by at least $10^{-4}$ over 15 epochs, restoring the weights from the epoch with the lowest validation loss. The model converged after approximately 91 epochs. Evaluation on the test set resulted in zero false positives and zero false negatives as expected for a cleaned dataset, confirming robust microlensing detection and satisfactory reconstruction quality.

\begin{figure}[!t]
    \centering
    \includegraphics[width=\columnwidth]{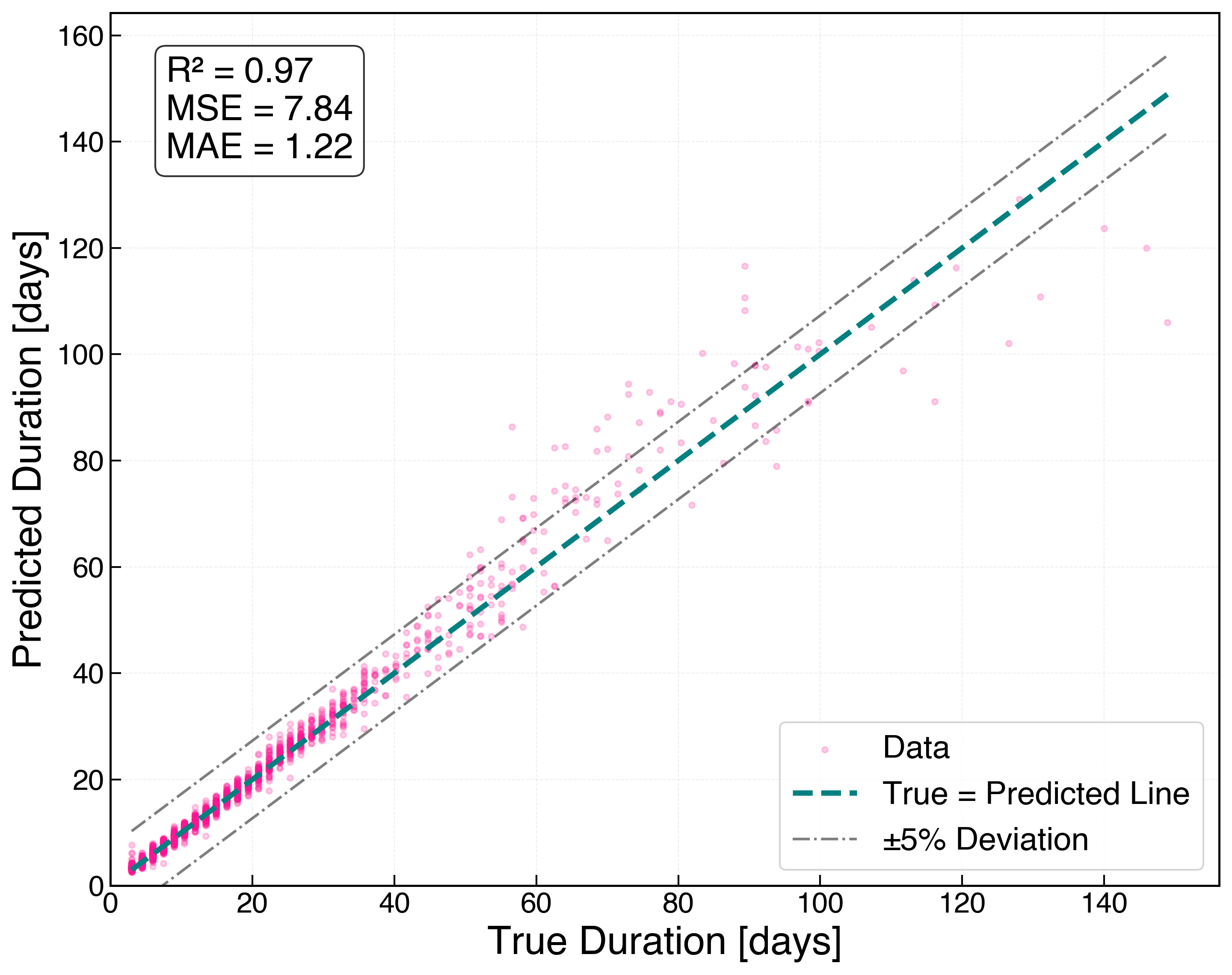}
    \caption{Predicted versus true microlensing event duration for all true-positive events in the test set. Microlensify achieves $R^{2} = 0.97$, MSE = 7.84, and MAE = 1.22. The teal dashed line represents the 1:1 relation, and the gray dash-dot lines indicate $\pm5\%$ deviation.}
    \label{fig:true_vs_predicted_duration}
\end{figure}

In the next step, we adjusted the loss weights to better balance the training objectives. The weights of the reconstruction and duration losses were increased to 10, while the KL term remained at 1.0, and the classification loss was increased to 50.0 to keep a strong focus on identifying microlensing events.

The model was then recompiled with the same optimizer and callback configurations and trained for up to 100 epochs. It converged after approximately 31 epochs, with the best weights (lowest validation loss) automatically restored. Evaluation on the test set again yielded zero false positives and zero false negatives, consistent with the use of a cleaned test set.
These weights and hyperparameters were chosen based on tests during development, which indicated improved reconstruction quality, more accurate event duration estimates, and a good classification performance.
Three true positive examples, including their light curves and the corresponding true and predicted duration values, are shown in \autoref{fig:true_positives_lightcurves}.
The scatter plot of predicted versus true event durations is shown in 
\autoref{fig:true_vs_predicted_duration}, achieving  
$R^2 = 0.97$, $\mathrm{MSE} = 7.84$, and $\mathrm{MAE} = 1.22$. 
The corresponding metrics are defined as
\begin{align}
R^{2} &= 1 - \frac{\sum_{i}(y_{i} - \hat{y}_{i})^{2}}{\sum_{i}(y_{i} - \bar{y})^{2}}, 
\label{eq:r2} \\[8pt]
\mathrm{MSE} &= \frac{1}{n}\sum_{i}(y_{i} - \hat{y}_{i})^{2}, 
&& \mathrm{MAE} &= \frac{1}{n}\sum_{i}|y_{i} - \hat{y}_{i}|,
\label{eq:mse_mae}
\end{align}
where $y_i$, $\hat{y}_i$, and $\bar{y}$ denote the true duration, the predicted duration, 
and the mean of the true durations, respectively.

\section{Results}\label{sec:results}
In this section, we explore Microlensify's predictions of single‑-lens microlensing events in TESS light curves extracted with different pipelines, as well as in light curves from other microlensing surveys, including OGLE, KMTNet, and MOA.

\subsection{Microlensify Predictions on Sector 12 of Eleanor}
\label{subsec:Microlensify on sec 12 Eleanor}

To evaluate the performance of our trained model, we applied it to approximately 
$1.5$ million TESS light curves from Sector~$12$ (no simulations included), sourced from the Eleanor pipeline (the same pipeline used for the training data), to identify potential microlensing events and estimate their durations and enable their reconstruction. Each light curve was standardized to a fixed length of $940$ data points to meet the input requirements of the model. For light curves with fewer than $940$ data points, padding was applied using values derived from the minimum flux, augmented with minor random variations ($0.1\%$ of the minimum flux), with padding positioned at the start or end based on the location of the minimum flux. For light curves exceeding $940$ data points, excess points were randomly removed to achieve the target length, then five statistics were computed for each light curve: minimum, maximum, median, standard deviation, and standard deviation divided by the difference between the maximum and minimum of the Eleanor\_Flux. The Eleanor\_Flux was normalized using the min--max method to ensure consistency with the scaling of the training data. The five features were log$_{10}$-transformed and then scaled using the per-feature scalers fitted on the training set. The resulting scaled feature vector, together with the normalized flux, was fed into the model to obtain a probability score for each light curve. Light curves achieving a probability of 
$1.0$ were classified as microlensing candidates, while others were categorized as non-microlensing. Furthermore, the duration of each identified microlensing event was estimated from the latent space of the model, which also enables event reconstruction.

\begin{table*}
\centering
\caption{Classification summary for TESS pipelines (Eleanor Sector 12, TESS-SPOC Sector 12, QLP Sector 12,          and QLP Sector 33).  
         Shown are the initial number of light curves processed, the number of sources assigned classification probability $P=1.0$ by the Microlensify model, and the final counts after quality cleaning ($\Delta\chi^2_{\rm KMTNet}>400$, peak SNR~$>30$, and position inside the skewness-–Von Neumann boundary with 0.1 tolerance).  
         The lower panel lists SIMBAD object types found in the cleaned high-confidence samples. Note that the object types are not deterministic, and the model prediction is based on microlensing-like peaks, which may be caused by microlensing false positives, blending with nearby sources, systematics, or asteroid crossings.}

\label{tab:simbad}
\normalsize                   
\setlength{\tabcolsep}{8pt}           
\renewcommand{\arraystretch}{1.25}       
\begin{tabular}{@{} l cccc @{}}
\toprule
 & Eleanor Sector12 & SPOC Sector12 & QLP Sector12 & QLP Sector33 \\
\midrule
Light curves fed to model & 1,500,000 & 133,329 & 2,857,972&  1,130,307 \\
Predicted with probability=1.0 & 545 & 108 & 54,091 &  11,770 \\
Remaining after threshold cleaning & 363 & 86 & 24,907 & 2,383 \\
Remaining after visual cleaning & 328 & 69 & 3,554 & 1,532 \\
\midrule
\multicolumn{4}{l}{SIMBAD Object Types in Final Clean Sample} \\[2pt]
Star & 6 & 19 & 317 & 238 \\
LongPeriodV* / Candidate & 30 & 25 & 1,975 & 195 \\
AGB$^*$ / Candidate & 0 & 2 & 9 & 35 \\
SB$^*$ (Spectroscopic Binary) & 0 & 0 & 68 & 18  \\
Variable$^*$ & 0 & 2 & 2 & 6  \\
S$^*$ / Candidate & 0 & 4 & 4 & 14  \\
HighPM$^*$ (High proper-motion star) & 0 & 7 & 8 & 10  \\
ChemPec* (Chemically Peculiar Star) & 0 & 1 & 20 & 7  \\
alf2CVnV* (alpha2 CVn Variable) & 0 & 1 & 0 & 1  \\
EmLine$^*$ (Emission-line Star) & 0 & 0 & 2 & 3  \\
C$^*$ (Carbon star) / Candidate & 2 & 1 & 5 & 4  \\
Mira & 1 & 0 & 52 & 3  \\
CataclyV* (Cataclysmic Variable)  & 0 & 1 & 1 & 1  \\
RGB$^*$ / Candidate & 0 & 0 & 1 & 3  \\
Eclipsing Binary / Candidate & 0 & 0 & 0 & 1  \\
Transient & 0 & 0 & 0 & 1  \\
Supergiant & 0 & 0 & 0 & 1  \\
Red Supergiant & 0 & 1 & 2 & 0  \\
Blue Supergiant & 0 & 0 & 1 & 0  \\
Symbiotic$^*$ & 0 & 0 & 1 & 0  \\
RRLyrae & 0 & 0 & 1 & 0  \\
Low-mass Star & 0 & 1 & 0 & 0  \\
YSO Candidate (Young Stellar Object) & 0 & 0 & 2 & 1  \\
HorBranch$^*$  (Horizontal Branch Star) & 0 & 0 & 0 & 1  \\
** (Double or Multiple Star) & 0 & 0 & 1 & 0  \\

\midrule
Not found in SIMBAD & 289 & 4 & 1,082 & 989  \\
\bottomrule
\end{tabular}
\end{table*}

\begin{figure*}
    \centering
    \includegraphics[width=\textwidth]{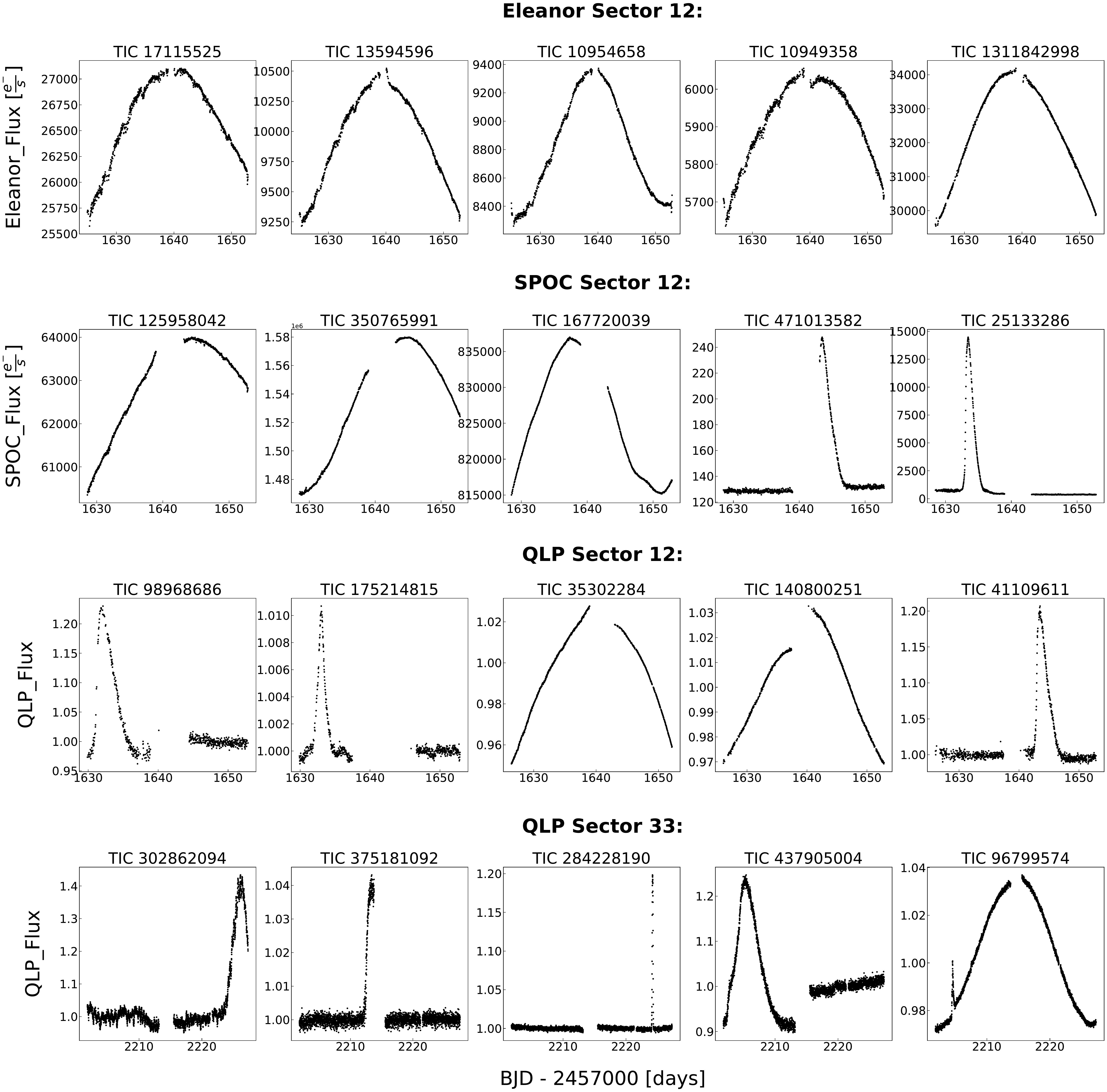}  
    \caption{Example light curves from the final high-confidence sample in each pipeline: Eleanor (Sector 12), SPOC (Sector 12), QLP (Sector 12), and QLP (Sector 33).  
         Each panel is labeled with its TIC ID.  
         Plotted flux is raw flux for Eleanor and SPOC pipelines, and normalized raw flux for QLP pipeline.  
         The x-axis shows time in days (BJD $-$ 2457000).  
         Corresponding sky coordinates (R.A., Dec), best-fit KMTNet parameters, SNR,  model-predicted 4FWHM values, and SIMBAD object types are reported in \autoref{tab:lightcurve_results}.}
    \label{fig:lightcurves_of_candidates}
\end{figure*}

\begin{table*}[!t]
\centering
\caption{Parameters for the example light curves shown in \autoref{fig:lightcurves_of_candidates}.  
         Listed are sky coordinates (R.A., Dec), best-fit KMTNet model parameters ($f_0$, $f_1$, $t_{\rm eff}$, $t_0$, $\Delta\chi^2_{\rm KMTNet}$), SNR, model-predicted 4FWHM values, and SIMBAD object types including long-period variables (LPV* and candidates), Mira variables, carbon stars (C*), EclBin (Eclipsing Binary), cataclysmic variables (CV*), asymptotic giant branch stars (AGB*), spectroscopic binaries (SB*), and regular stars; a fraction of sources have no counterpart in SIMBAD ("Not Found"). The flux parameters $f_0$ and $f_1$ are given in units of $e^{-}\,\mathrm{s}^{-1}$ for Eleanor and SPOC light curves and are dimensionless for QLP light curves.
}
\label{tab:lightcurve_results}
\renewcommand{\arraystretch}{1.4}
\small
\setlength{\tabcolsep}{3pt}
\begin{tabular}{@{} c c c c c c c c c c c l @{}}
\toprule
TIC ID & Mission & RA [$^\circ$]  & DEC [$^\circ$] & $f_0$ & $f_1$ & $t_{\rm eff}$ [days]& $t_0$ (days) & $\Delta \chi^2$ & SNR & $\mathrm{4FWHM}_{\rm pred}$ [days] & Obj Type \\
\midrule
17115525     & Eleanor 12 & 247.8453 & -53.7273 & 22837.4  & 4288.4   & 14.25 & 1639.22 & 990.9  & 103.4 & 95.8 & LPV* Cand \\
13594596     & Eleanor 12 & 247.4081 & -53.4260 & 6283.6   & 4149.7   & 14.25 & 1639.22 & 1055.9 & 38.5  & 52.8 & LPV* Cand \\
10954658     & Eleanor 12 & 247.2055 & -52.9932 & 7591.2   & 1817.9   & 6.012  & 1639.001 & 1053.5 & 38.0  & 30.5 & LPV* Cand \\
10949358     & Eleanor 12 & 247.1984 & -52.2926 & 4892.5   & 1160.3   & 14.250 & 1639.224 & 975.6  & 79.2  & 86.1 & Mira \\
1311842998   & Eleanor 12 & 259.6461 & -63.9124 & 18875.4  & 15232.7  & 14.250 & 1639.223 & 1025.8 & 35.5  & 67.4 & Not Found \\
125958042    & SPOC 12    & 256.8225 & -46.7657 & 43881.7  & 19999.7  & 25.333 & 1645.488 & 912.5  & 65.7  & 77.0 & C* \\
350765991    & SPOC 12    & 89.2000 & -59.4225 & 1376695.6& 208844.5 & 8.015  & 1644.520 & 906.9  & 42.1  & 44.2 & LPV* \\
167720039    & SPOC 12    & 103.4345 & -65.9139 & 797226.4  & 40692.8   & 6.012  & 1638.509 & 898.1  & 118.1 & 26.3 & EclBin \\
471013582    & SPOC 12    & 91.1287  & -71.4231 & $-$118.1 & 135.0    & 0.619  & 1643.598 & 931.6  & 300.6 & 5.1  & Not Found \\
25133286     & SPOC 12    & 62.2975  & -71.2949 & $-$15.9  & 15832.9  & 0.299  & 1633.686 & 679.1  & 251.5 & 4.4  & CV* \\

98968686 & QLP 12 & 236.2427 & -32.4584 & 0.990 & 0.290 & 0.421 & 1632.279 & 573.3 & 34.7 & 3.70 & Not Found \\
175214815 & QLP 12 & 235.1873 & -41.6747 & 0.999 & 0.012 & 0.237 & 1633.051 & 597.2 & 57.7 & 3.0 & Star \\
35302284 & QLP 12 & 249.790 & -52.849 & 0.531 & -0.369 & 13.303 & 1639.735 & 885.5 & 48.3 & 35.6 & AGB* \\
140800251 & QLP 12 & 74.7482 & -72.5115 & 0.883 & 0.144 & 9.977 & 1639.377 & 868.7 & 58.3 & 48.2 & LPV* \\
41109611 & QLP 12 & 91.1188 & -71.4320 & 0.979 & 0.239 & 0.421 & 1643.772 & 886.1 & 73.9 & 4.5 & Star \\
302862094    & QLP 33    & 94.6352  & 8.7629 & 0.969      & 0.491  & 0.619 & 2226.320 & 3185.8 & 77.6  & 4.7 & Not Found  \\
375181092    & QLP 33     & 133.5796 & -76.2657 & 1.0      & 0.1      & 0.299  & 2213.413 & 3340.2 & 297.4 & 5.1  & Not Found \\
284228190    & QLP 33     & 108.7522 & 4.0401     & 0.5      & 0.5      & 0.100  & 2224.300 & 2927.7 & 41.1  & 3.1  & Star \\
437905004    & QLP 33     & 94.1609  & 15.4007    & 1.0      & 0.3      & 0.619  & 2205.392 & 2837.1 & 41.5  & 8.2  & Not Found \\
96799574     & QLP 33     & 105.1177 & -30.7597 & 0.9      & 0.1      & 8.015  & 2215.192 & 3255.5 & 46.8  & 30.3 & SB* \\
\bottomrule
\end{tabular}
\end{table*}

From the $1.5$ million light curves processed, the model identified $545$ microlensing candidates, corresponding to a detection rate of 0.0363\%. Note that the entire Eleanor Sector~$12$ contains $\sim$$19.5$ million light curves.

In the next step, we applied additional cleaning criteria, as explained in \autoref{sec:Methodology}, to retain only the most probable microlensing candidates. These criteria include $\Delta\chi^{2}_{\mathrm{KMTNet}} > 400$, SNR $> 30$, and placement in the skewness--von Neumann ratio space with a 
tolerance of $10^{-1}$. The results of this cleaning procedure are summarized in \autoref{tab:simbad}. For the Eleanor predictions, $182$ 
of the $545$ model candidates were removed.
We also visually inspected all remaining candidates after the automated cleaning stage. For the Eleanor Sector-12 sample, $35$ of the $363$ candidates were discarded, leaving $328$ high-quality events.

We then searched the remaining candidates within a $1''$ radius in SIMBAD\footnote[11]{\url{https://simbad.u-strasbg.fr/simbad/sim-fbasic}} to identify their object types, as listed in \autoref{tab:simbad}. The object types are not deterministic, as the model detects microlensing--like peaks that can arise from false positives, blending with nearby sources, or systematics.
 Examples of the different object types found in the Eleanor Sector~12 dataset are shown in \autoref{fig:lightcurves_of_candidates}, and their KMTNet fit parameters, SNR values, and the $4\,$FWHM duration estimates predicted by the model are listed in \autoref{tab:lightcurve_results}.

\subsection{Cross-matching Gaia Alerts with TESS}
\label{subsec:Cross-matching Gaia Alerts with TESS}
\begin{figure*}[t]
    \centering
    \includegraphics[width=\textwidth]{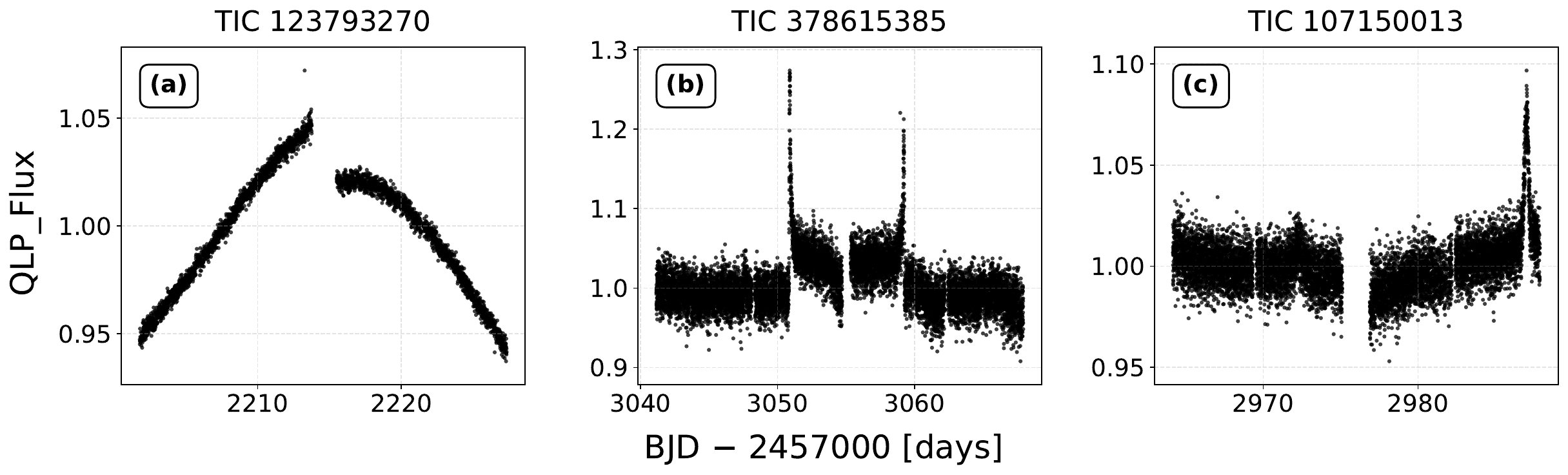}
    \caption{Light curves of three microlensing candidates (all extracted from the QLP pipeline).
(a) TIC 123793270, cross-matched with Gaia20fnr, observed in TESS Sector 33; Microlensify predicts probability 1.0.
(b) TIC 378615385, cross-matched with Gaia23bra, observed in Sector 64; a clear binary microlensing event well extracted by QLP but not classified as microlensing by our model.
(c) Free-floating planet candidate reported by \citet{Kunimoto2025} in Sector 61; Microlensify successfully recovers it as a microlensing candidate.}
    \label{fig:three_candidates}
\end{figure*}

To provide an overview of the different types of microlensing events that TESS can detect, and to test its ability to capture such events despite its large pixel size of 21 arcseconds, we cross--matched Gaia Alerts with TESS observations. This cross--match was performed before the publication of the papers reporting the events discussed below. We find that Gaia20fnr\footnote[9]{\url{https://gsaweb.ast.cam.ac.uk/alerts/alert/Gaia20fnr/}}  
was observed under TIC ID~$123793270$ in Sector~33, with its light curve extracted only by the QLP pipeline, as shown in \autoref{fig:three_candidates}a. This event is a binary-‑lens microlensing event, as analyzed in detail by \citet{wicker2026gaia20fnr}, but the part observed by TESS corresponds to the wings of the light curve, which resemble a single‑-lens event. Thus, we would expect Microlensify to be able to detect it. 

As explained in \autoref{subsec:ffis}, the QLP pipeline normalizes each light curve individually for each $13.7$-day orbit. Consequently, raw flux is not available and within each $27.4$-day observing period, the two orbits are normalized separately. Despite these limitations, we tested our model's performance on QLP data to determine whether it could recover this Gaia candidate. The QLP light curves in Sector~$33$ 
have a $10$-minute cadence, whereas our model was trained on $30$-minute-cadence 
data from Sector~$12$. To make predictions robust and applicable regardless of cadence, 
we implemented the following additional preprocessing pipeline.
For each light curve, we considered multiple window sizes $N$, ranging from 1000 points to half the total number of points in the light curve in steps of 1000 points. 
For a given window size $N$, the light curve was divided into non-overlapping contiguous chunks of exactly $N$ points each; any remaining points after the last full chunk were added to the final chunk.
Each chunk was then uniformly downsampled to exactly 1000 points by selecting every 
$(N // 1000)$-th point. 
From these 1000 points, we randomly subsampled 940 points to produce a fixed-length input segment matching the model's expected shape.
For the specific case of $N = 1000$, adding the remainder to the final chunk can make it substantially longer than 1000 points (e.g., up to nearly 2000 points, depending on the light curve length). Randomly subsampling such an extended chunk down to 940 points risks eliminating short-duration events located in that final region. In contrast, for larger window sizes ($N > 1000$), these same short events are reliably captured by the shorter chunks processed earlier in the pipeline. Thus, only when $N = 1000$, we generate one additional prediction using the terminal 1000 contiguous points of the light curve (then reduced to 940 points by random subsampling).

Additionally, a global prediction was obtained by downsampling the entire light curve in the same way (step = total length // 1000) and randomly reducing it to 940 points, thereby providing an overview of the entire light curve.
We then applied min--max normalization to the down--sampled QLP\_FLUX values (which are normalized) to rescale them to the training range $[0,1]$. 
Since raw flux is not provided by QLP, the five statistical features were assigned the median values derived from our training set including both microlensing and non-microlensing data: minimum flux = $189.912$, maximum flux = $396.818$, median flux = $310.555$, standard deviation = $7.381$, and 
$\mathrm{std}/(\mathrm{max}-\mathrm{min}) = 0.0357$. These values were log$_{10}$-transformed and normalized using the same per-feature scalers fitted on the training data. The resulting 940-point normalized flux segment, together with the five scaled statistical features, was fed to the model, which produced a classification probability and a predicted 4FWHM for each segment. 

It is important to note that the predicted 4FWHM is returned by the model in units of the 27.4-day TESS sector duration used during training. To obtain the physical duration in days, we simply rescale the model output by the actual time span covered by the 940-point input segment: we multiply the predicted value by (segment time span in days / 27.4). Because the true 4FWHM values in the training set ranged from approximately 2.98 to 166.8 days, the rescaled physical 4FWHM is most reliable when the 940 points cover a baseline close to one full sector ($\sim$27.4 days). For much shorter or longer segments, the reported 4FWHM should be treated as an approximate indicator rather than a precise measurement.

Across the entire $1{,}130{,}307$ light curves of QLP Sector~$33$ processed under these conditions, the model identified $11{,}770$ candidates with probability $1.0$, corresponding to a detection rate of 1.041\%, successfully including the known microlensing event Gaia20fnr. 
We then applied the same cleaning criteria ($\Delta\chi^{2}_{\mathrm{KMTNet}} > 400$, SNR $> 30$, and placement in the skewness–-von Neumann ratio space with a tolerance of $10^{-1}$) to obtain the most probable microlensing candidates and removed 9,387 of the 11,770 predicted candidates. We also visually checked the remaining candidates, and 851 of the 2,383 surviving candidates were removed, leaving 1,532 events, which is a good result given the limitations of the QLP pipeline (lack of raw flux and use of the fixed set of five statistical features derived from the training set). Despite this limitation of the QLP pipeline, the procedure was efficient: from the initial 1,130,307 light curves in QLP Sector~33, only 2,383 candidates (∼0.2\%) required human inspection.

We then searched the remaining candidates, as done previously for Eleanor Sector~12, within a $1''$ radius in SIMBAD to determine their object types, as listed in \autoref{tab:simbad}. 
Some examples of the different object types identified in the QLP Sector 33 dataset are shown in \autoref{fig:lightcurves_of_candidates}, and their KMTNet fit parameters, SNR values, and the $4\,\mathrm{FWHM}$ duration estimates predicted by the model are listed in \autoref{tab:lightcurve_results}. One notable case is TIC 284228190, classified as a star in SIMBAD, whose light curve shows self‑-lensing–-like pulses that are in fact caused by an asteroid crossing, representing a false positive for such events \citep{Sorabella2023}.

We further note that cross-matching found Gaia23bra\footnote[10]{\url{https://gsaweb.ast.cam.ac.uk/alerts/alert/Gaia23bra/}} in Sector~$64$ of the QLP (TIC ID $378615385$; independently identified and analyzed by \citet{harris2026tess}), which exhibits a caustic-crossing signature and is not a simple PSPL event. As expected, our model trained primarily on single--lens events did not assign probability $1.0$ to this object. Nevertheless, its clear detection in the QLP light curve (\autoref{fig:three_candidates}b) serves as strong evidence of the QLP pipeline's capability to reliably capture complex microlensing events in its raw data.

\subsection{Microlensify Prediction for an Individual FFP Candidate}
\label{subsec:Microlensify Prediction for an Individual FFP Candidate}

In a recent study, \citet{Kunimoto2025} searched for free-‑floating planet microlensing events in 7.5 million light curves from TESS Sectors 61–-65, using a Box-‑Least‑-Squares (BLS) algorithm adapted to detect single, nonperiodic brightening events with durations between 17 minutes and 1 day. They find a short‑-duration microlensing event candidate whose light curve morphology is consistent with that expected from a low‑-mass free‑-floating planet (FFP), although it appears inconsistent with current estimates of FFP abundance in this mass range, raising the possibility that the signal could instead be a false positive such as a stellar flare, heartbeat binary, or centrifugal breakout.

This event was observed in the 200-second-cadence data of Sector 61 of QLP pipeline. When processed individually with Microlensify (using the same chunking, preprocessing, and the fixed set of five statistical features optimized for the QLP pipeline as discussed in 
\autoref{subsec:Cross-matching Gaia Alerts with TESS}) our model classified this event as a microlensing event with probability 1.0 (\autoref{fig:three_candidates}c).

\subsection{Microlensify Predictions on SPOC and QLP Sector 12}
\label{subsec:Microlensify Predictions on SPOC and QLP Sector 12}

The Microlensify training set is based on Eleanor light curves from TESS Sector 12. Here, we evaluate the model's ability to make predictions on light curves extracted by the SPOC and QLP pipelines for the same sector.

The SPOC pipeline provides a more highly selective set of target stars, as described in \autoref{subsec:ffis}. For Sector 12, the SPOC dataset contains about $133{,}329$ extracted light curves. We applied the same preprocessing procedure used for the Eleanor data, but using the SPOC\_Flux. From these light curves, the model identified $108$ microlensing candidates, corresponding to a detection rate of $0.0810\%$. We then applied the same additional cleaning criteria used for the Eleanor dataset ($\Delta\chi^{2}_{\mathrm{KMTNet}} > 400$, SNR $> 30$, and placement in the skewness--von Neumann ratio space with a tolerance of $10^{-1}$) to select the most probable microlensing candidates. After this step, $22$ of the $108$ candidates were discarded. We also visually checked all remaining candidates after the automated cleaning stage. For the SPOC Sector 12 sample, $17$ of the $86$ remaining candidates were rejected, leaving $69$ events.
We then cross-matched the candidates with SIMBAD within a $1''$ radius and identified several object types as listed \autoref{tab:simbad}.

Examples of the different object types identified in the SPOC Sector 12 dataset are shown in \autoref{fig:lightcurves_of_candidates}, and their KMTNet fit parameters, SNR values, and the $4\,\mathrm{FWHM}$ duration estimates predicted by the model are listed in \autoref{tab:lightcurve_results}.

\begin{figure*}[t!]
    \centering
    \includegraphics[width=\textwidth]{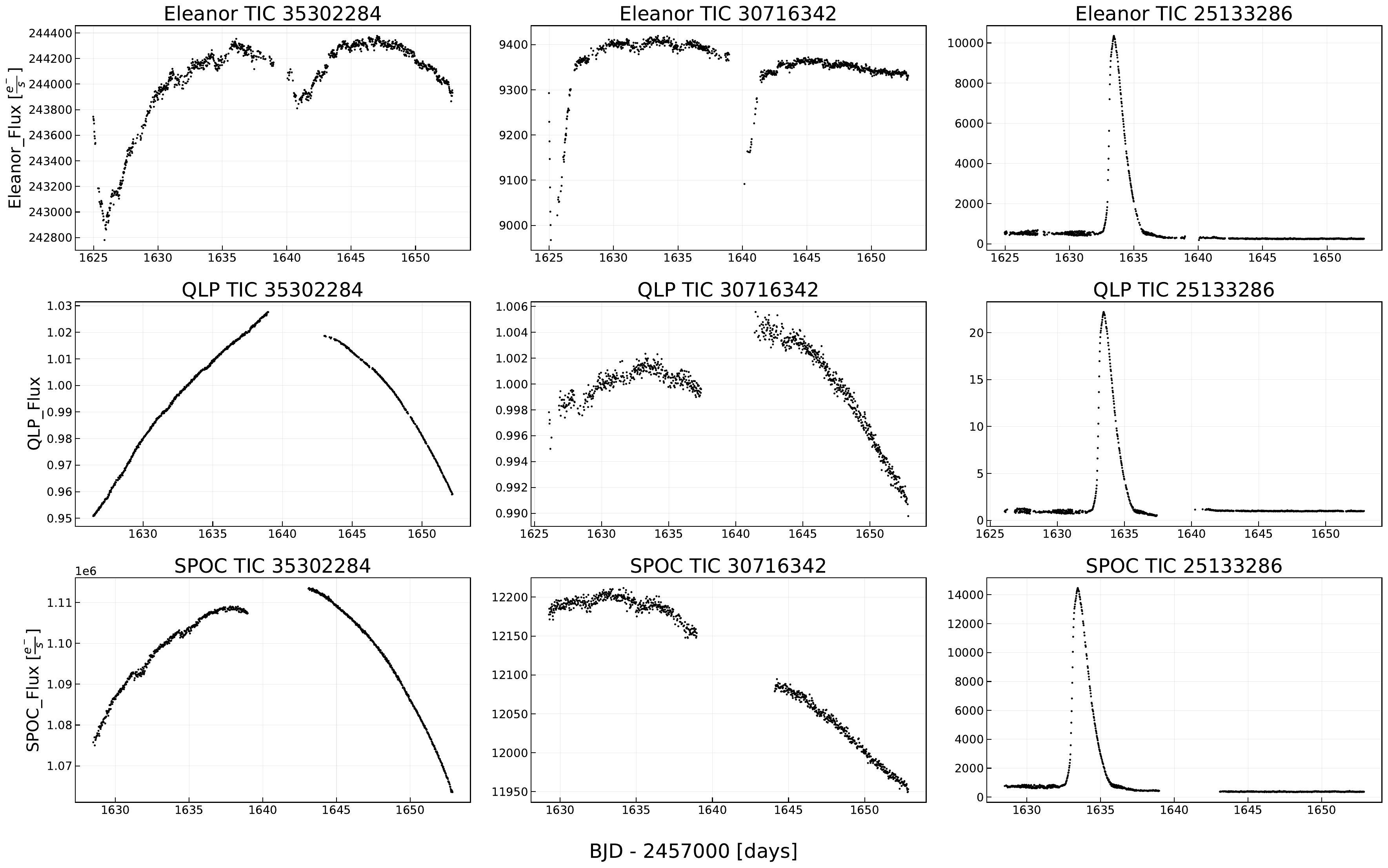}
    \caption{Three examples from TESS Sector 12 processed by all three light curve extraction pipelines (Eleanor, QLP, and SPOC). 
TIC 35302284 (SIMBAD type: AGB star): Microlensify probability 1.0 in QLP and SPOC, poor-quality Eleanor light curve. 
TIC 30716342 (SIMBAD type: star): clear normalization offset across the sector gap in QLP, cleanly extracted by SPOC, poor Eleanor extraction; not selected as a microlensing candidate in any pipeline. 
TIC 25133286 (SIMBAD type: cataclysmic variable): good-quality extraction and probability 1.0 in all three pipelines.}
    \label{fig:compare_pipelines}
\end{figure*}

As discussed in \autoref{subsec:Cross-matching Gaia Alerts with TESS}, the QLP pipeline does not provide raw flux, and within each $27.4$-day observing period the two $13.7$-day orbits are normalized independently. Because raw flux is unavailable, the five statistical features were assigned the median values derived from our training set. We then applied a min--max normalization to the QLP\_Flux values (which are already independently normalized) to rescale them to the training range $[0,1]$, and fed them to Microlensify.
The full QLP Sector~12 dataset contains $2{,}857{,}972$ light curves. Microlensify labeled $54{,}091$ of them as microlensing candidates with a class probability of 1.0. We then applied the same cleaning metrics used for the Eleanor and SPOC samples ($\Delta\chi^{2}_{\mathrm{KMTNet}} > 400$, SNR $> 30$, and placement in the skewness--von~Neumann ratio space with a tolerance of $10^{-1}$), reducing the sample to $24{,}907$ candidates. We also visually checked the remaining candidates and kept 3{,}554 of them. Most of the removed ones are microlensing-like systematics. The absence of raw flux in QLP, combined with the fact that our training set was based on Eleanor light curves, where such systematics were not labeled as non-microlensing, resulted in the removal of 21{,}353 candidates. Nevertheless, only about $0.87\%$ of the entire QLP Sector~12 dataset required human verification, a reasonable result given these limitations.

The cross-matching results of the final QLP Sector~12 candidates with SIMBAD within a $1''$ radius are listed in \autoref{tab:simbad}. Examples of these object types are shown in \autoref{fig:lightcurves_of_candidates}, and their fit parameters and duration estimates are listed in \autoref{tab:lightcurve_results}.

Many of the sources detected in Sector~12 appear in all three pipelines, Eleanor, SPOC, and QLP, allowing direct comparison of their extracted light curves.

 In \autoref{fig:compare_pipelines}, we show three targets from Sector 12 that were processed with all three pipelines for direct comparison of the three light curve extraction methods. TIC 35302284 is assigned probability 1.0 in both QLP and TESS-SPOC, but the Eleanor light curve is of poor quality. This object is classified as an asymptotic giant branch (AGB) star in SIMBAD. TIC 30716342 shows an obvious normalization offset across the sector gap in QLP, whereas TESS-SPOC extracts it cleanly. The Eleanor extraction is again poor. This target is listed as a star in SIMBAD and is not selected as a candidate by our model in any pipeline. TIC 25133286 is well extracted in all three pipelines, receives probability 1.0 in each, and is classified as a cataclysmic variable in SIMBAD.

Among all candidates cross-‑matched with SIMBAD, we found several single‑-lens false positives, such as long‑-period variables, cataclysmic variables, AGB stars, Mira variables, RGB stars, and supergiants. We also found sharp, self‑-lensing–-like peaks that are actually caused by asteroid crossings, such as TIC 284228190 shown in \autoref{fig:lightcurves_of_candidates}. The final microlensing candidates are therefore the objects classified as stars and those not found in SIMBAD, which require further analysis.

\subsection{Model Predictions on Microlensing Candidates from Other Surveys}
\label{subsec:Prediction on others}

As a final check of the model's ability to generalize, we applied it to real microlensing events discovered by the three main ground-based surveys: OGLE, KMTNet, and MOA.

We randomly selected 200 published events from each survey, covering many different observing seasons. The light curves were downloaded from the public archives: OGLE\footnote[12]{\url{https://ogle.astrouw.edu.pl/ogle4/ews/ews.html}}, KMTNet\footnote[13]{\url{https://kmtnet.kasi.re.kr/ulens/}}, and MOA\footnote[14]{\url{https://moaprime.massey.ac.nz/moaarchive/}}. Magnitudes were converted to relative flux. 

We then applied exactly the same preprocessing and prediction pipeline designed to allow Microlensify to generalize across different cadences, as discussed in 
\autoref{subsec:Cross-matching Gaia Alerts with TESS}. This includes applying quality cuts, multi-window chunking (1000 points up to the full length), uniform downsampling to 1000 points, random selection of 940 points, applying per-segment min–-max normalization to the flux values, computing the five scalar features from the segment itself, log--transforming these features and scaling them using the scalers fitted on the training set, and finally generating the Microlensify predictions.

The results were remarkably consistent across the three surveys:

\begin{itemize}
    \item OGLE: 186/200 with probability $\ge 0.99$
    \item KMTNet (CTIO $I$-band): 184/200 with probability $\ge 0.99$
    \item MOA: 186/200 with probability $\ge 0.99$
\end{itemize}

This corresponds to an average detection efficiency of 92.66\,\% for events with classification probability of $\ge 0.99$.

The small number of events that fell below this threshold almost always exhibited obvious problems: very sparse sampling across the peak, almost no baseline coverage, or the presence of high-flux outlier points near the event that confuse the model. When the light curve around the peak is reasonably well sampled and includes a visible baseline, the model recognises the event essentially without fail.

Regarding the event duration estimation, since the true 4FWHM values in the training set range approximately from 2.98 to 166.8 days, the rescaled physical 4FWHM is most reliable when derived from predictions on the entire light curve downsampled to 940 points, rather than from predictions on chunks in different parts of the light curve. Therefore, for short-duration events that are not detected over the whole light curve window,  the duration estimation is less trustworthy.

These tests show that a classifier trained solely on a single sector of 30-minute-cadence TESS data can reliably identify microlensing events in completely independent ground-based surveys with very different cadence, photometric precision, and passband.

\section{Summary and Conclusions}
\label{sec:concl}

In this work, we have developed Microlensify, a Physics-Informed Transformer-Based Variational Autoencoder (PI-TVAE) designed specifically for the detection of single-lens microlensing events in light curves extracted from TESS full-frame images. In addition to classification, Microlensify reconstructs the light curve and estimates the microlensing event duration.

The model was trained on approximately 17,000 simulated microlensing events and a similar number (\(\sim\)17,500) of real non-microlensing light curves from Sector 12 extracted from Eleanor pipeline. 
Each input consists of a fixed-length sequence of 940 normalized flux with five statistical features: minimum, maximum, median, standard deviation, and standard deviation divided by the difference between the maximum. 
To handle light curves of arbitrary length, we implement a chunking procedure that splits longer time series into overlapping 940-point segments, as detailed in \autoref{subsec:Cross-matching Gaia Alerts with TESS}. 
Microlensify reconstructs light curves and predicts the event duration with $R^2 = 0.97$.

We tested the trained Microlensify model on  Sector 12 light curves from three different TESS pipelines: On approximately 1.5 million Eleanor light curves, the model flagged $545$ objects with classification probability 1.0. On the stricter SPOC FFI-extracted sample ($133{,}329$ light curves), it recovered $108$ candidates. On the QLP catalogue (approximately 2.86 million light curves from the same sector), the model yielded $54{,}091$ high-probability candidates (using fixed statistical features, since QLP does not provide raw flux).
After applying traditional cleaning thresholds ($\Delta\chi^2_{\rm KMTNet} > 400$, peak SNR $> 30$, and location within the skewness--von Neumann ratio boundary with tolerance $10^{-1}$) followed by visual inspection, the final numbers of high-quality candidates were:
$328$ (Eleanor Sector 12), 
$69$ (TESS-SPOC Sector 12), 
and $3{,}554$ (QLP Sector 12).

We removed $21{,}353$ QLP sector 12 light curves during visual inspection. This is mainly because QLP Sector 12 contains microlensing‑-like systematics that were not included as non-‑microlensing examples in our training set, which was built from Eleanor Sector 12 light curves. In addition, since QLP does not provide raw flux, we could not compute the five statistical features directly from the raw flux. Instead, we used fixed values for these features based on their median values in the training set. These limitations together explain the relatively large number of rejected QLP candidates. Nevertheless, only about $0.87\%$ of the entire QLP Sector 12 dataset required human verification.

We also cross-matched known microlensing events from the Gaia Photometric Science Alerts with TESS targets.
Among them, Gaia20fnr (Sector 33, \autoref{fig:three_candidates}a) was recovered by Microlensify with classification probability 1.0.
Across the entire QLP Sector 33 ($1{,}130{,}307$ light curves), the model initially assigned probability 1.0 to $11{,}770$ targets. After applying the cleaning criteria and visually inspecting the remaining candidates, $1{,}532$
light curves remained, representing $0.2\%$ of the QLP Sector 33 dataset that required human verification.

The known binary--lens event Gaia23bra (\autoref{fig:three_candidates}b) was not assigned high confidence, which is expected given that Microlensify was trained exclusively on single--lens simulations; nevertheless, a clear flux variation is visible in the TESS pixels, confirming that even TESS's large 21 arcseconds pixels can capture significant signal from binary--lens events.
Additionally, the candidate free-floating planet event reported by \citet{Kunimoto2025} (\autoref{fig:three_candidates}c) was also flagged by our pipeline with probability 1.0, further validating the model's sensitivity to short--time scales.

These final candidates were cross-matched with the SIMBAD database listed in \autoref{tab:simbad} including stars, long--period variables, AGB stars, spectroscopic binaries, variable stars, S--type stars, high proper--motion stars, chemically peculiar stars, $\alpha^{2}$CVn variables, emission line stars, carbon stars, Mira variables, cataclysmic variables, RGB stars, eclipsing binaries, a transient (the class assigned to Gaia20fnr under TIC 123793270 in TESS, which Microlensify correctly identified as a microlensing event), supergiants, red supergiants, a blue supergiant, a symbiotic star, an RR~Lyrae, young stellar objects, a classical Cepheid Variable, a double or multiple star system. Our final microlensing candidates are sources classified as stars or those that have not yet been classified, which require further analysis to confirm whether they represent real microlensing signals. In addition, we identified several sharp peaks in Sector 33 of the QLP data that are consistent with self--lensing events but are actually caused by asteroid crossings \citep{Sorabella2023}, representing an important source of false positives for high--cadence surveys such as Roman.

The diversity of false-positive types is noticeably greater in the QLP sample than in the Eleanor and TESS-SPOC samples. 
Eleanor light curves preserve raw flux but are dominated by systematic noise. 
QLP effectively removes most systematic noise in the light curve, yet it lacks raw flux and applies independent normalization to each ~13.7-day orbit, which often distorts long-timescale signals across the mid-sector gap. 
The TESS-SPOC FFI extraction, by contrast, applies strict brightness and contamination criteria, resulting in far fewer light curves than are produced by Eleanor or QLP.

We also evaluated Microlensify on published microlensing events from OGLE, KMTNet, and MOA surveys. 
The model achieves an average detection efficiency of 92.7\% at classification probability $\geq 0.99$, making it fully competitive with traditional search pipelines. The event duration estimates are reliable only when derived from predictions on the full light curve downsampled to the 940‑-point format used during training; duration predictions made on shorter light curve segments are less trustworthy.

In terms of computational speed, Microlensify classifies a 940-point light curve in $\sim$0.1\,s on a single core of an Apple M1 Pro CPU roughly  300 times faster than the complete KMTNet event-detection and fitting pipeline executed on the same hardware, making it ideal for initial screening of millions of targets before deeper vetting.

In contrast to other recent models that train only on simulated data to obtain clean true/false labels, we used real high--cadence TESS light curves to include real false positives of microlensing events in high--cadence surveys.

For future work, we will further investigate the candidates from our cleaned lists, particularly those with good baseline coverage, and perform precise PSPL fits to validate the signals.
Although the current model was trained on segments of only 940 data points, the chunking procedure we implemented allows it to make robust predictions on light curves of different lengths without any limitation. Nevertheless, we plan to make it more precise by training on longer sequences and on multi-sector light curves (those targets that have been observed in more than one sector).
Additionally, we plan to train microlensify on Roman simulations and extract other microlensing parameters in addition to the event duration.

Microlensify is publicly available as an open-source package on GitHub\footnote{\url{https://github.com/Atousa-Kalantari/Microlensify}}, allowing the community to use it as an initial search tool for single--lens microlensing events in time--domain surveys.

\section{ACKNOWLEDGEMENTS}\label{sec:ACKNOWLEDGEMENTS}

We warmly thank the TESS Help Desk team, in particular Nicole~E.~Schanche, Rebekah~Hounsell, and Tyler~Pritchard, for their assistance with numerous TESS-related questions. We are grateful to Michelle~Kunimoto, Chelsea~Huang, Glen~Petitpas, and Jack~Haviland for guidance on the QLP pipeline; to Brian~Powell, Adina~Feinstein, and Benjamin~Montet for help with the Eleanor pipeline; and to Douglas~Caldwell for detailed explanations regarding the TESS-SPOC pipeline. We thank Parisa~Sangtarash for clarifying aspects of microlensing light curve simulation, Jennifer~Yee, Hongjing~Yang, and Qiyue~Qian for advice on KMTNet data and algorithms, Radek~Poleski for assistance with OGLE data, and Stela~Ishitani Silva for help with MOA data. We also thank Andrew~Vanderburg and Vikram~Bhamre for helpful discussions about this work.

This paper includes data collected by the TESS mission, publicly available from the Mikulski Archive for Space Telescopes (MAST). Funding for the TESS mission is provided by NASA's Science Mission Directorate. We have used public TESS light curves produced by the TESS-SPOC pipeline \citep{Caldwell2020} (\url{https://archive.stsci.edu/hlsp/tess-spoc}), the GSFC-Eleanor-Lite pipeline \citep{Powell2022} (\url{https://archive.stsci.edu/hlsp/gsfc-eleanor-lite}), and the Quick-Look Pipeline (QLP) (\citet{Huang2020a}; \citet{Huang2020b}; \citet{Kunimoto2021}; \citet{Kunimoto2022})(\url{https://archive.stsci.edu/hlsp/qlp}).

This work has also made use of data from the Korean Microlensing Telescope Network 
(KMTNet; \url{https://kmtnet.kasi.re.kr/ulens/}), 
the Microlensing Observations in Astrophysics collaboration 
(MOA; \url{https://it019909.massey.ac.nz/moa/}), 
the Optical Gravitational Lensing Experiment 
(OGLE; \url{http://ogle.astrouw.edu.pl}), 
and the Gaia Alerts service operated for the European Space Agency mission Gaia 
(\url{https://gsaweb.ast.cam.ac.uk/alerts}).

Facilities: TESS, KMTNet, MOA, OGLE, Gaia.

This research has made use of the following software packages: \texttt{TensorFlow} \citep{Abadi2016}, \texttt{numpy} \citep{harris2020array}, \texttt{pandas} \citep{reback2020pandas}, \texttt{astropy} \citep{astropy:2013,astropy:2018,astropy:2022}, \texttt{scipy} \citep{2020SciPy-NMeth}, \texttt{lightkurve} \citep{2018ascl.soft12013L}, \texttt{Eleanor} \citep{Feinstein2019, ascl:1905.007}, and \texttt{tess--point} \citep{2020ascl.soft03001B}. 

This research has made use of the SIMBAD database, operated at CDS, Strasbourg, France \citep{2000A&AS..143....9W}.
Predictions were performed with our open-source package \texttt{Microlensify}, publicly available on GitHub at \url{https://github.com/Atousa-Kalantari/Microlensify}. All additional scripts and notebooks used in this work are provided in the same repository.

\bibliography{cite}
\bibliographystyle{aasjournal}

\end{document}